\documentclass{article}
\usepackage[utf8]{inputenc}
\usepackage{fullpage}
\usepackage{enumitem}
\usepackage{booktabs}
\usepackage{multirow}
\usepackage{natbib}
\usepackage{subcaption}
\usepackage{graphicx}
\usepackage{amsmath}
\usepackage{amsfonts}
\usepackage{amssymb}
\usepackage{nicefrac}
\usepackage{xcolor}
\usepackage{hyperref}
\usepackage{csquotes}
\usepackage{mdframed}
\usepackage{tablefootnote}
\usepackage{authblk}
\usepackage{float}
\usepackage{placeins}
\usepackage{authblk}

\newcommand{\cmu}{a certain university}

\newcommand{\constant}{\textsc{Trivial}}
\newcommand{\tpms}{\textsc{TPMS}}
\newcommand{\elmo}{\textsc{ELMo}}
\newcommand{\specter}{\textsc{Specter}}
\newcommand{\mfr}{\textsc{Specter+MFR}}
\newcommand{\acl}{\textsc{ACL}}
\newcommand{\weblink}{\url{https://forms.gle/SP1Rh8eivGz54xR37}}
\newcommand{\openai}{\textsc{o1-mini}}
\newcommand{\claude}{\textsc{Claude Sonnet 3.5}}
\newcommand{\gemini}{\textsc{Gemini 2 Flash}}
\newcommand{\spectertwo}{\textsc{Specter2}}

\title{A Gold Standard Dataset for the Reviewer Assignment Problem}
\author{Ivan Stelmakh}
\author{John Wieting}
\author{Sarina Xi}
\author{Graham Neubig}
\author{Nihar B. Shah\thanks{Corresponding author: \href{mailto:nihars@cs.cmu.edu}{nihars@cs.cmu.edu}}}

\affil{\textit{Carnegie Mellon University}}

\date{}

\begin{document}

\maketitle

\begin{abstract}
    Many peer-review venues are either using or looking to use algorithms to assign submissions to reviewers. The crux of such automated approaches is the notion of the ``\emph{similarity score}''---a numerical estimate of the expertise of a reviewer in reviewing a paper---and many algorithms have been proposed to compute these scores. However, these algorithms have not been subjected to a principled comparison, making it difficult for stakeholders to choose the algorithm in an evidence-based manner. 
    The key challenge in comparing existing algorithms and developing better algorithms is the lack of the publicly available gold-standard data that would be needed to perform reproducible research. 
    {We address this challenge by collecting a novel dataset of similarity scores} that we release to the research community. Our dataset consists of 477 self-reported expertise scores provided by 58 researchers who evaluated their expertise in reviewing papers they have read previously. 
    
    We use this data to compare several popular algorithms currently employed in computer science conferences and come up with recommendations for stakeholders. Our four main findings are: 
    \begin{itemize}[leftmargin=20pt, itemsep=0pt, topsep=1pt]
        \item All algorithms make a non-trivial amount of error. For the task of ordering two papers in terms of their relevance for a reviewer, the error rates range from 12\%-30\% in easy cases to 36\%-43\% in hard cases, thereby highlighting the vital need for more research on the similarity-computation problem.
        
        \item Most specialized algorithms are designed to work with titles and abstracts of papers, and in this regime the \spectertwo{} algorithm performs best.

        \item The classical TF-IDF algorithm which can use full texts of papers is on par with \spectertwo{} that uses only titles and abstracts.

        \item The performance of off-the-shelf LLMs is worse than the specialized algorithms.
    \end{itemize} 
We encourage researchers to participate in our survey and contribute their data to the dataset.
\end{abstract}

\section{Introduction}

Assigning papers to reviewers with appropriate expertise is the key prerequisite for high-quality reviews \citep{thurner2011peer, black1998makes, bianchi2015three}. Even a small fraction of incorrect reviews can negatively impact the quality of the published scientific standard~\citep{thurner2011peer} and hurt careers of researchers~\citep{merton1968matthew, squazzoni2012saint, thorngate2014numbers}. Quoting~\citet{triggle07future}:
\begin{quote}
    \emph{``An incompetent review may lead to the rejection of the submitted paper, or of the grant application, and the ultimate failure of the career of the author.''}
\end{quote}

Conventionally, the selection of reviewers for a submission was the task of a journal editor or a program committee of a conference. However, the rapid growth in the number of submissions to many publication venues~\citep{shah2021survey} has made the manual selection of reviewers extremely challenging. As a result, many peer-review venues in computer science as well as other fields are either using or looking to use algorithms to assist organizers in assigning submissions to reviewers~\citep{Garg2010papers, charlin13tpms, kobren19localfairness, kerzendorf2020distributed, stelmakh21pr4a, OpenReview2022Similarities}.

The key component of existing automated approaches for assigning reviewers to submissions is the ``\textit{similarity score}''. For any reviewer-submission pair, the similarity score is a number that captures the expertise of the reviewer in reviewing the submission. The assignment process involves first computing the similarity score for each submission-reviewer pair, and then using these scores to assign the reviewers to submissions~\cite[Section 3]{shah2021survey}. The similarity scores are typically computed by matching the text of the submission with the profile (e.g., past papers) of the reviewer, which is the primary focus of this paper. Additionally, these scores may also be augmented by manually selected subject areas or reviewer-provided manual preferences (``bids'').

Several algorithms for computing similarity scores have been already proposed and used in practice (we review these algorithms in Sections~\ref{section:relatedwork} and~\ref{section:expsetup}). However, there are two key challenges in designing and using such algorithms in a principled manner. First, there is no publicly available gold standard data that can be used for algorithm development. Second, despite the existence of many similarity-computation algorithms, the absence of the gold standard data prevents principled comparison of these algorithms. As a result, three flagship machine learning conferences---ICML, NeurIPS, and ACL---rely on three different similarity-computation algorithms, and the differences in performance of these algorithms are not well understood. 

In this work, \emph{we address the aforementioned challenges and collect a dataset of reviewers' expertise that can facilitate the progress in the reviewer assignment problem}. Specifically, we conduct a survey of computer science researchers whose experience level ranges from graduate students to senior professors. In the survey, we ask participants to report their expertise in reviewing computer science papers they read over the last year. 

\paragraph{Contributions} 
First, we collect and release a high-quality dataset of reviewers' expertise that can be used for training and/or evaluation of similarity-computation algorithms. The dataset can be found on the project's GitHub page:\vspace{-7pt}
    \begin{center}
\url{https://github.com/niharshah/goldstandard-reviewer-paper-match}
    \end{center}    
    
We use the collected dataset to compare existing similarity-computation algorithms in terms of their accuracy in predicting the similarities, and inform organizers in making a principled choice for their venue. Specifically, we observe that when all algorithms operate with titles and abstracts of papers, the most advanced \spectertwo{} algorithm performs best. However, when the much simpler  \tpms{} is additionally provided with full texts of papers, it achieves the same level of performance. Generic large language models do not perform as well. We also conduct additional studies pertaining to `pooling' methods and the use of titles+abstracts versus full papers. 

\smallskip

Overall, we observe that all current algorithms exhibit non-trivial amounts of error. When tasked to compare a pair of papers in terms of the expertise of a given reviewer, all algorithms make 12\%-30\% mistakes even when two papers are selected to have a large difference in reviewer's expertise. On a more challenging task of comparing two high-expertise papers, the algorithms err with probability 36\%-43\%. This observation underscores the vital need to design better algorithms for matching reviewers to papers.

\paragraph{Key limitations and a call for participation} It is important to keep in mind the following limitations. The first limitation is its size, in that it comprises 477 data points contributed by 58 researchers. Second, there is a prominent bias in terms of geographical skew with 74\% of participants from the US. 

Our dataset is not set in stone and we encourage researchers to participate in our survey and contribute their data to the dataset. By collecting more samples, we enable more fine-grained comparisons and also improve the diversity of the dataset in terms of both population of participants and subject areas of papers. Any contributions herein will benefit the peer-review process.  The survey is available at:\vspace{-3pt}
\begin{center}
    \weblink 
\end{center}
and we will be updating the released version regularly.

\section{Related literature}
\label{section:relatedwork}

In this section, we discuss relevant past studies. We begin with an overview of works that report comparisons of different similarity-computation algorithms. We then provide a brief discussion of the commonly used procedure for computing similarity scores, which are ultimately utilized to assign reviewers to submissions. Finally, we conclude with a list of works that design algorithms to automate other aspects of reviewer assignment.

\paragraph{Evaluation of similarity-computation algorithms} The OpenReview platform~\citep{OpenReview2022Similarities} uses an approach of predicting authorship as a proxy to measuring the quality of the similarity scores computed by any algorithm. Specifically, they consider papers authored by a number of researchers, remove one of these papers from the corpus, and predict expertise of each researcher in reviewing the selected paper. The performance of an algorithm then is measured as a fraction of times the author of the selected paper is predicted to be among the top reviewers for this paper. This authorship proxy, however, may not be representative of the actual task of similarity computation as algorithms that accurately predict the authorship relationship (and hence do well according to this approach) are not guaranteed to accurately estimate expertise in reviewing submissions authored by other researchers.

\citet{mimno07topicbased} collected a dataset with external expertise judgments. Specifically, they used 148 papers accepted to the NeurIPS 2006 conference and 364 reviewers from the NeurIPS 2005 conference and ask human annotators -- independent established researchers -- to evaluate expertise for a selected subset of 650 (paper, reviewer) pairs. Similarly, \citet{zhao2022reviewer} collected the set of papers and reviewers in the ICIP 2016 conference. They then asked independent researchers to provide their estimates of the match between various reviewer-paper pairs. While this approach results in a publicly available dataset, we note that external expertise judgments may also be noisy as judges may have incomplete information about the expertise of reviewers.

\citet{dumais1992automating}, \citet{rodriguez08coauthorsip} and \citet{anjum19pare} obtain more accurate expertise judgments by relying on self reports of expertise evaluations from reviewers. In more detail, \citet{dumais1992automating} and \citet{rodriguez08coauthorsip} rely on \emph{ex-ante} bids---preferences of reviewers in reviewing submissions made in advance of reviewing. In contrast, \citet{anjum19pare} rely on \emph{ex-post} evaluations of expertise made by reviewers after reviewing the submissions. These works construct datasets that can be used to evaluate algorithms: \citet{dumais1992automating} employ 117 papers and 15 reviewers, \citet{rodriguez08coauthorsip} employ 102 papers and 69 reviewers and \citet{anjum19pare} employ 20 papers and 33 reviewers. However, \citet{rodriguez08coauthorsip} and \citet{anjum19pare} use sensitive data that cannot be released without compromising the privacy of reviewers. Furthermore, \emph{ex-ante} evaluations of~\citet{dumais1992automating} and~\citet{rodriguez08coauthorsip} may not have a high accuracy as (i) bids may contaminate expertise judgments with \emph{willingness} to review submissions, and (ii) bids are based on a very brief acquaintance with papers. On the other hand, \emph{ex-post} data by~\citet{anjum19pare} is collected for papers assigned to reviewers using a specific similarity-computation algorithm. Thus, while collected evaluations have high precision, they may also have low recall if the employed similarity-computation algorithm erroneously assigned low expertise scores to some (paper, reviewer) pairs as evaluations of expertise for such papers were not observed.

In this work, we collect a novel dataset of reviewer expertise that (i) can be released publicly, and (ii) contains accurate self-evaluations of expertise that are based on a deep understanding of the paper and are not biased towards any existing similarity-computation algorithm.

\paragraph{Similarity scores in conferences} In modern conferences, similarity scores are typically computed by combining two types of input:
\begin{itemize}[leftmargin=20pt, itemsep=0pt, topsep=5pt]
    \item \emph{Initial automated estimates.} First, a similarity-computation algorithm is used to compute initial estimates. Many algorithms have been proposed for this task~\citep{mimno07topicbased, rodriguez08coauthorsip, charlin13tpms, liu14graphpropagation, tran17expertsuggestion, anjum19pare,kerzendorf2020distributed, OpenReview2022Similarities} and we provide more details on several algorithms used in flagship computer science conferences in Section~\ref{section:expsetup}.

    \item \emph{Human corrections.} Second, automated estimates are corrected by reviewers who can read abstracts of submissions and report bids---preferences in reviewing the submissions. A number of works focus on the bidding stage and (i) explore the optimal strategy to assist reviewers in navigating the pool of thousands of submissions~\citep{fiez2019super, meir2020market} or (ii) protect the system from strategic bids made by colluding reviewers willing to get assigned to each other's paper~\citep{jecmen2020manipulation, ruihan21making,boehmer2021combating,jecmen22dataset}.
\end{itemize}

Combining these two types of input in a principled manner is a non-trivial task. As a result, different conferences use different strategies~\citep{shah2017design,leyton2022matching} and there is a lack of empirical or theoretical evidence that would guide venue organizers in their decisions.

\paragraph{Automation of the assignment stage} At a high level, automated assignment stage consists of two steps: first, similarity scores are computed; second, reviewers are allocated to submissions such that some notion of assignment quality (formulated in terms of the similarity scores) is maximized. In this work, we focus on the first step of the process. However, for completeness, we now mention several works that design algorithms for the second step.

A popular notion of assignment quality is the cumulative similarity score, that is, the sum of the similarity scores across all assigned reviewers and papers. An algorithm pursuing such an objective is implemented in the widely employed TPMS assignment algorithm~\citep{charlin13tpms} and similar ideas are explored in many papers~\citep{goldsmith07aiconf, tang10constraied, Long13gooadandfair}. While the cumulative objective is a natural choice, it has been noted that it may discriminate against certain submissions by allocating all irrelevant reviewers to a subset of submissions, even when a more balanced assignment exists~\citep{Garg2010papers}. Thus, a number of works has explored the idea of assignment fairness, aiming at producing more balanced assignments~\citep{kobren19localfairness, stelmakh21pr4a}. Finally, other works explore the ideas of envy-freeness~\citep{tan21envy, payan22fair}, resistance to lone-wolf strategic behavior~\citep{xu2018strategyproof,dhull2022price}, and encouraging various types of diversity~\citep{Li15concert,leyton2022matching}.

\section{Data collection pipeline}
\label{section:pipeline}

In this section, we describe the process of data collection. This work was performed under the approval of an Institutional Review Board (IRB).

\paragraph{Gold standard data} In this study, we aim to collect a dataset of self-evaluations of reviewers' expertise that satisfies two desiderata:
\begin{enumerate}[leftmargin=25pt, itemsep=0pt, topsep=1pt, label={(D\arabic*})]
    \item The dataset should comprise evaluations of expertise for papers participants read to a reasonable extent.\label{desiderata:one}
    \item  The dataset should be released publicly without disclosing any sensitive information.\label{desiderata:two}
    \item We want the dataset to comprise experts' ratings of their own expertise rather than people judging other peoples' expertise.
\end{enumerate}
Now, there are a few natural approaches one could consider to collect a dataset. For instance, one could consider data from the bidding process which may be publicly released (or at least a subset of it~\citealp{Shah_2022_randomizedtransparency}) but this will violate desideratum 1. An approach to satisfy desideratum 1 would be to observe that reviewers do read assigned papers at a conference and then release that data; however, one cannot release information about who reviewed which papers and hence this approach cannot satisfy desideratum 2. In order to satisfy both 1 and 2, we could ask researchers to provide their opinions on the goodness of match between a set of papers and potential reviewers~\citep{mimno07topicbased}, but this will violate desideratum 3 and may also violate desideratum 1. 

We thus took an approach in this paper that satisfies all three desiderata: asking researchers to provide their expertise levels for papers they have read. More specifically, researchers are asked to provide a list of papers they have read and provide a rating/ranking of these papers according to their own expertise for reviewing them.  We now discuss our data collection procedure in more detail. 

\paragraph{Participant recruiting} We recruited participants using a combination of several channels that are typically employed to recruit reviewers for computer science conferences:
\begin{itemize}[leftmargin=20pt, itemsep=0pt, topsep=5pt]
    \item \textit{Mailing lists.} First, we sent recruiting emails to relevant mailing lists of several universities and research departments of companies.

    \item \textit{Social media.} Second, via a call for participation on X (formerly Twitter). 
    
    \item \textit{Personal communication.} Third, we sent personal invites to researchers from our professional network.
\end{itemize}

We had a screening criterion requiring that prospective participants have at least one paper published in the broad area of computer science. Overall, for the version of the dataset we release in this paper, we managed to recruit 58 participants, all of whom passed the screening.

\paragraph{Expertise evaluations} The key idea of our approach to expertise evaluation is to ask participants to \emph{evaluate their expertise in reviewing papers they read in the past}. After reading a paper, a researcher is in the best possible position to evaluate whether they have the right background---both in terms of the techniques used in the paper and in terms of the broader research area of the paper---to judge the quality of the paper. With this motivation, we asked participants to:
\begin{quote}
    \textit{Recall 5-10 papers in their broad research area that they read to a reasonable extent in the last year and tell us their expertise in reviewing these papers.}
\end{quote}
In more detail, the choice of papers was constrained by two minor conditions:
\begin{itemize}[leftmargin=20pt, itemsep=0pt, topsep=5pt]
    \item The papers reported by a participant should not be authored by them.
    \item The papers reported by a participant should be freely available online.
\end{itemize}

In addition to these constraints, we gave several recommendations to the participants in order to make the dataset more diverse and useful for the research purposes:

\begin{itemize}[leftmargin=20pt, itemsep=0pt, topsep=5pt]
    \item First, we asked participants to choose papers that cover the whole spectrum of their expertise with some papers being well-separated (e.g., very high expertise and very low expertise) and some papers being nearly-tied (e.g., two high-expertise papers).
    
    \item Second, we recommended participants to avoid ties in their evaluations. To help participants comply with this recommendation, we implemented evaluation on a scale of 1 (``I am not qualified to review this paper'') to 5 (``I have background necessary to evaluate all the aspects of the paper'') with a 0.25 step size, enabling participants to report papers with small differences in expertise.

    \item Third, we asked participants to come up with papers that they think may be tricky for existing similarity-computation algorithms. For this, we relied on the commonsense understanding and did not instruct participants on the inner-workings of these algorithms. We only provided an example: \emph{a naive computation of similarity may think that a paper on ``Theory of Information Dissemination in Social Networks'' has high similarity with an Information Theory researcher, but in reality, this researcher may not have expertise in reviewing this paper.}
\end{itemize}

Overall, the time needed for participants to contribute to the dataset was estimated to be 5--10 minutes. The full instructions of the survey are available in Appendix~\ref{appendix:survey}.

\paragraph{Data release} Following the procedure outlined above, we collected responses from 58 researchers. These responses constitute an initial version of the dataset that we release in this work. Each entry in the dataset corresponds to a participant and comprises evaluations of their expertise in reviewing papers of their choice. For each paper and each participant, we provide representations that are sufficient to start working on our dataset:

\begin{itemize}[itemsep=0pt, leftmargin=20pt, topsep=5pt]
    \item \textit{Participant.} Each participant is represented by their Semantic Scholar ID and complete bibliography crawled from Semantic Scholar on May 1, 2022.
    
    \item \textit{Paper.} Each paper, including papers from participants' bibliographies, is represented by its Semantic Scholar ID, title, abstract, list of authors, publication year, and arXiv identifier. Additionally, papers from participants' responses are supplied with links to freely available PDFs (whenever available).
\end{itemize}

\section{Data exploration}
\label{section:exploration}

In this section we explore the collected data and present various characteristics of the dataset. The subsequent sections then detail the results of using this data to benchmark various popular similarity-computation algorithms.

\subsection{Participants} 
\label{section:exploration:participants}

We begin with a discussion of the pool of the survey participants: Table~\ref{table:participants_demo} displays its key characteristics. First, we note that all participants work in the broad area of computer science and have rich publication profiles (at least two papers published, with the mean of 54 papers and the median of 20 papers). In many subareas of computer science, including machine learning and artificial intelligence, having two papers is usually sufficient to join the reviewer pool of flagship conferences. Given that approximately 85\% of participants either have PhD or are in the process of getting the degree, we conclude that most of the researchers who contributed to our dataset are eligible to review for machine learning and artificial intelligence conferences.

Second, we caveat that most of the participants are male researchers affiliated with US-based organizations, with about 40\% of all participants being affiliated with \cmu. Hence, the population of participants is not necessarily representative of the general population of the machine learning and computer science communities. We encourage researchers to be aware of this limitation when using our dataset. We note that the data collection process does not conclude with the publication of the present paper and we will be updating the dataset as new responses come.  We also encourage readers to contribute 5--10 minutes of their time to fill out the survey \weblink{} and make the dataset more representative.

\begin{table}[t]
\begin{center}
\textsc{Total number of participants: 58} \medskip \\
\begin{small}
\begin{sc}
\begin{tabular}{llr}
\toprule
 Characteristic &  Quantity & Value \\
\midrule
Gender & \% Male & 78 \\ 
\midrule
Country & \% USA & 74 \\
\midrule
\multirow{3}{*}{Position} & \% PhD student & 45 \\
 & \% Faculty &  28 \\
 & \% Post-PhD (non-faculty) &  12  \\
\midrule
\multirow{4}{*}{Experience} & Min \# publications & 2 \\
 & Max \# publications & 492 \\
 & Mean \# publications & 54 \\
 & Median \# publications & 20 \\

\bottomrule
\end{tabular}
\end{sc}
\end{small}
\end{center}
\caption{Demography of participants. For the first four characteristics, quantities represent percentages of the one or more most popular classes in the dataset. Note that classification is done manually based on publicly available information and may not be free of error. For the last characteristic, quantities are computed based on Semantic Scholar profiles.}
\label{table:participants_demo}
\end{table}

\subsection{Papers}
\label{section:exploration:papers}

We now describe the set of papers that constitute our dataset. Overall, participants evaluated their expertise in reviewing 463 unique papers. Out of these 463 papers, 12 papers appeared in reports of two participants, 1 paper was mentioned by three participants, and the remaining papers were mentioned by one participant each.

Table~\ref{table:papers_demo} presents several characteristics of the pool of papers in our dataset. First, we note that all but one of the papers are listed on Semantic Scholar, enabling similarity-computation algorithms developed on the dataset to query additional information about the papers from the Semantic Scholar database. For the paper that does not have a Semantic Scholar profile, we construct such a profile manually and keep this paper in the dataset.  Additionally, most of the papers (99\%) have their PDFs freely available online, thereby allowing algorithms to use full texts of papers to compute similarities.

Semantic Scholar has a built-in tool to identify broad research areas of papers~\citep{wade2022semantic}. According to this tool, 99\% of the papers included to our dataset belong to the broad area of computer science---the target area for our data-collection procedure. The remaining four papers belong to the neighboring fields of mathematics, philosophy, and the computational branch of biology. Finally, approximately 75\% of all papers in our dataset are published in or after 2020, ensuring that our dataset contains recent papers that similarity-computation algorithms encounter in practice.

\begin{table}[t]
\begin{center}
\textsc{Total number of papers: 463} \medskip \\
\begin{small}
\begin{sc}
\begin{tabular}{llr}
\toprule
 Characteristic &  Quantity & Value \\
\midrule
\multirow{3}{*}{Open Access} & \# On semantic scholar & 462 \\ 
& \# On arXiv & 411 \\ 
& \# PDF available & 457 \\ 
\midrule
Research Areas & \# Computer science & 459 \\
\midrule
\multirow{2}{*}{Publication Year} & \% Before 2020 & 25 \\
& \% 2020 or later & 75 \\
\bottomrule
\end{tabular}
\end{sc}
\end{small}
\end{center}
\caption{Characteristics of the 463 papers in the released dataset. Most of the papers are freely available online and belong to the broad area of computer science.}
\label{table:papers_demo}
\end{table}

\subsection{Evaluations of expertise}

\begin{figure*}[b]
\centering
\begin{subfigure}{.6\textwidth}
  \centering
  \includegraphics[width=\linewidth]{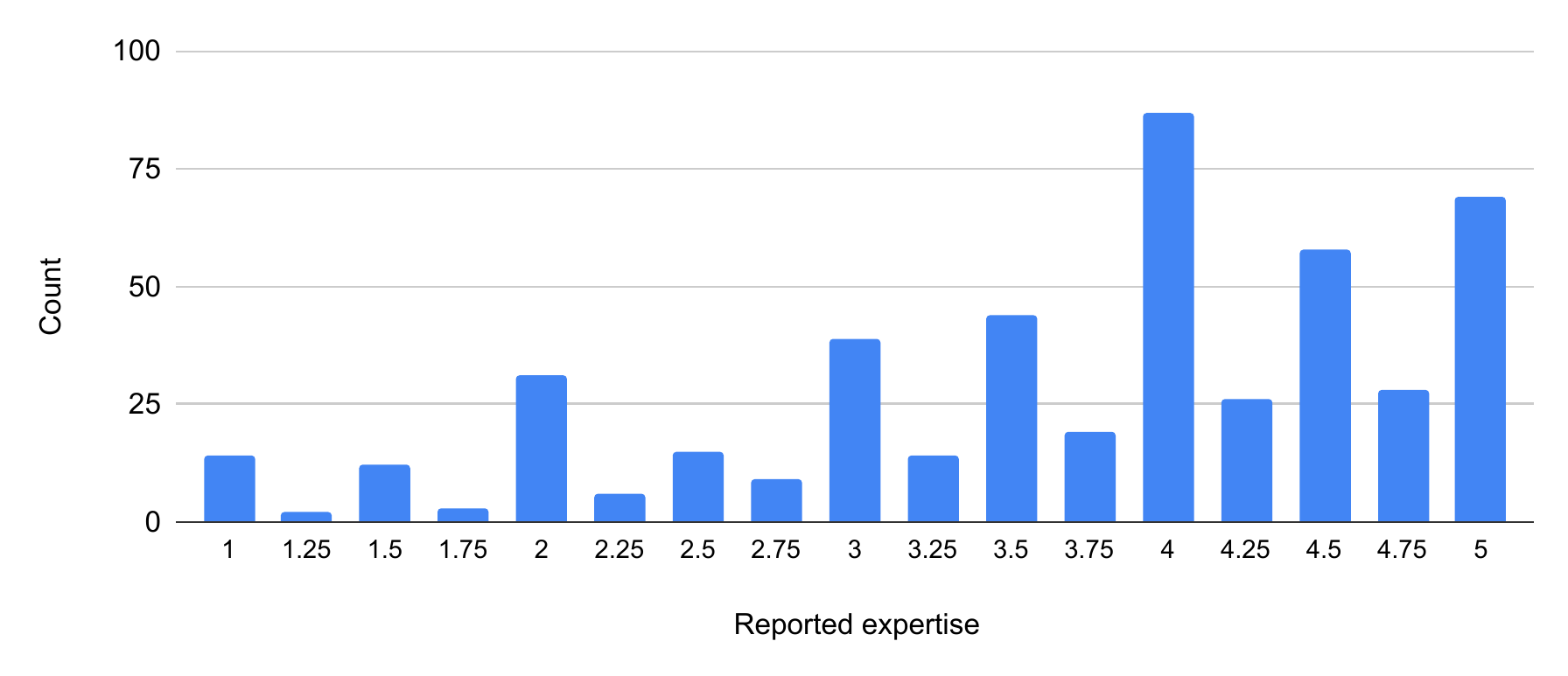}
  \caption{Counts of expertise values reported by participants.}
  \label{fig:experise:marginal}
\end{subfigure}\hfill%
\begin{subfigure}{.4\textwidth}
  \centering
  \includegraphics[width=0.8\linewidth]{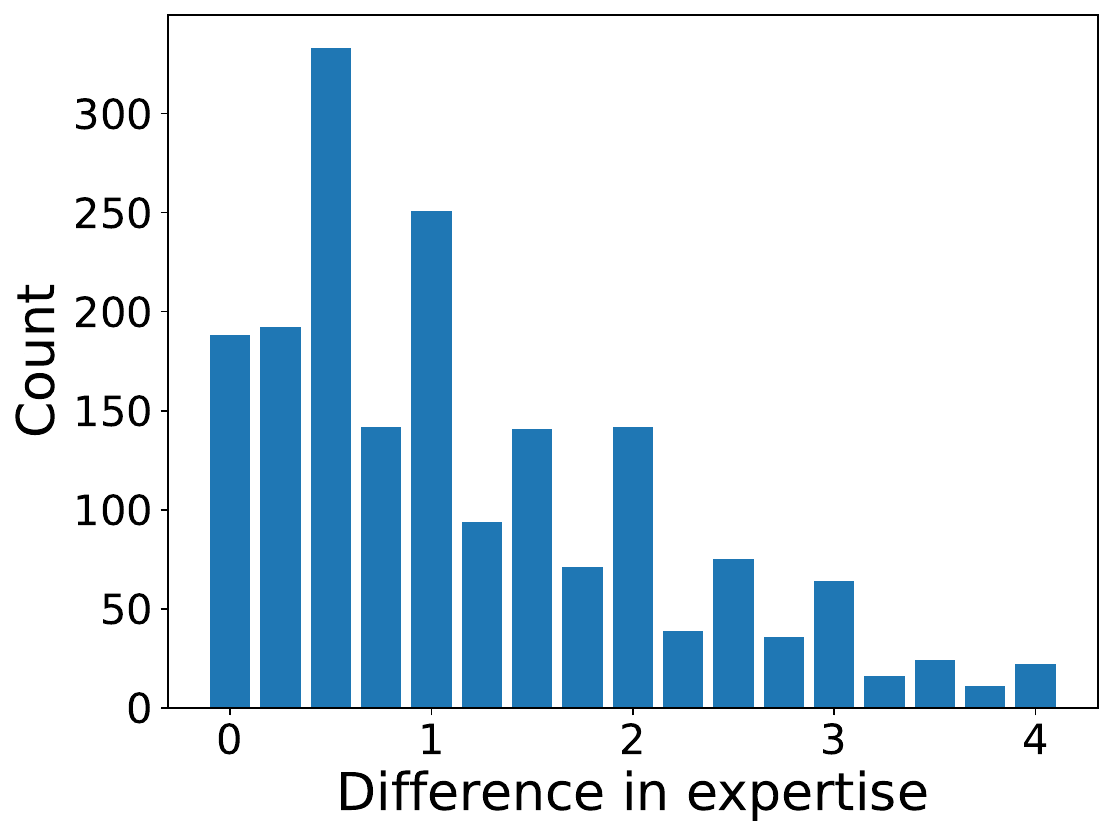}
  \caption{Counts of differences in expertise evaluations.}
  \label{fig:experise:diff}
\end{subfigure}
\caption{Distribution of expertise scores reported by participants.}
\label{fig:experise}
\end{figure*}

Finally, we proceed to the key aspect of our dataset---evaluations of expertise in reviewing the papers reported by participants. All but one participant reported expertise in reviewing at least 5 papers with the mean number of papers per participant being 8.2 and the total number of expertise evaluations being 477.\footnote{To clarify, the dataset does \emph{not} contain the evaluations by all participants for each of the 463 unique papers; the data only consistes of the evaluations provided by each participant on the set of papers reported by that participant, making it a total of 477 evaluations.}

Figure~\ref{fig:experise} provides visualization of expertise evaluations made by participants. First, Figure~\ref{fig:experise:marginal} displays the counts of expertise values. About 56\% of the reported papers have an expertise score greater than equal to 4, thereby yielding a good spread. 

Second, Figure~\ref{fig:experise:diff} shows the distributions of pairwise differences in expertise evaluations made by the same reviewer. To construct this figure, for each participant we considered all pairs of papers in their report. Next, we pooled the absolute values of the pairwise differences in expertise across participants. We then plotted the histogram of these differences in the figure. Observe that the distribution in Figure~\ref{fig:experise:diff} has a heavy tail, suggesting that our dataset is suitable for evaluating the accuracy of similarity-computation methods both at a coarse level (large differences between the values of expertise) and at a fine level (small differences between the values of expertise).

\section{Experimental setup}
\label{section:expsetup}

We now describe the setup of experiments on our dataset.

\subsection{Metric}
\label{section:setup:metric}

\newcommand{\participant}{r}
\newcommand{\paperset}{\mathcal{P}}
\newcommand{\paper}{p}
\newcommand{\expertiseset}{\mathcal{E}}
\newcommand{\expertise}{\varepsilon}
\newcommand{\similarityset}{\mathcal{S}}
\newcommand{\similarity}{s}
\newcommand{\numpapers}{m}
\newcommand{\loss}{L}
\newcommand{\indicator}{\mathbb{I}}

We begin with defining a main metric that we use in this work to evaluate performance of the algorithms. For this, we rely on the Kendall's Tau distance that is closely related to the widely used Kendall's Tau~\citep{kendall1938new} rank correlation coefficient. We introduce the metric and the algorithms in this section, followed by the results in Section~\ref{section:results}. 
Subsequently in Section~\ref{section:section:easyhard}, we provide a stratified evaluation that separates out hard and easy instances. Finally in Appendix~\ref{section:metrics} we evaluate a number of additional metrics. 

\paragraph{Intuition} Before we introduce the metric in full details, let us provide some intuition behind it. Consider a pair of papers and a participant who evaluated their expertise in reviewing these papers. We say that a similarity-computation algorithm makes an error if it fails to correctly predict the paper for which the participant reported the higher value of expertise.

Of course, some pairs of papers are harder to resolve than others (e.g., it is harder to resolve papers with expertise scores 4.0 and 4.25 than 4.0 and 1.0). To capture this intuition, whenever an algorithm makes an error, we penalize it by \emph{the absolute difference in the expertise values reported by the participant}. Overall, the loss of the algorithm is the sum of losses across all pairs of papers evaluated by participants normalized to take values between 0 and 1 (the lower the better).

\paragraph{Formal definition} The intuition behind the metric that we have just introduced should be sufficient to interpret the results of our study so the reader may skip the rest of this section and move directly to Section~\ref{section:setup:algorithms}. However, for the sake of rigor, we now introduce our metric more formally.

Consider any algorithm that produces real-valued predictions of reviewers' expertise in reviewing a given set of papers. We call these predictions the ``similarity scores'' given by the algorithm. We assume that a higher value of a similarity score means that the algorithm  predicts a better expertise. The metric we define below is agnostic to the range or exact values of predicted similarity scores; it only relies on the relative values across different papers.

Now consider any participant $\participant$ in our study. We let $\numpapers_\participant$ denote the number of papers for which participant $\participant$ reports their expertise. We let $\paperset_{\participant} = \{\paper_{\participant}^{(1)}, \paper_{\participant}^{(2)}, \ldots, \paper_{\participant}^{(\numpapers_\participant)}\}$ denote this set of $\numpapers_\participant$ papers. For every $i \in \{1,\ldots,\numpapers_\participant\}$, we let $\expertise_{\participant}^{(i)}  \in \left\{1, 1.25, 1.5, \ldots, 5\right\}$ denote the expertise self-reported by participant $\participant$ for paper $\paper_{\participant}^{(i)}$. Next, for every $i \in \{1,\ldots,\numpapers_\participant\}$, we let $\similarity_{\participant}^{(i)}$ denote the real-valued similarity score given by the algorithm to the pair (reviewer $\participant$, paper $\paper_{\participant}^{(i)}$). 

Having set up this notation, we now define the `unnormalized' loss of the algorithm with respect to participant $\participant$ as:

\begin{align*}
    \loss_{\participant} \! = \! &\sum\limits_{\substack{i, j = 1 \\ i < j}}^{\numpapers_\participant} \! \Bigg( \! \underbrace{\indicator \left\{ (\similarity_{\participant}^{(i)} \! - \similarity_{\participant}^{(j)}) \! \times \! (\expertise_{\participant}^{(i)} \! - \expertise_{\participant}^{(j)}) < 0 \right\}}_{\text{error}} \! \times \big| \expertise_{\participant}^{(i)} \! - \expertise_{\participant}^{(j)} \big| + \underbrace{\indicator \left\{ (\similarity_{\participant}^{(i)} \! - \similarity_{\participant}^{(j)}) \! \times \! (\expertise_{\participant}^{(i)} \! - \expertise_{\participant}^{(j)}) = 0 \right\}}_{\text{tie}} \! \times  \frac{1}{2} \big|  \expertise_{\participant}^{(i)} - \expertise_{\participant}^{(j)}  \big| \! \Bigg).
\end{align*}

In words, for each pair of papers $(\paper_{\participant}^{(i)}, \paper_{\participant}^{(j)})$ reported by participant $\participant$, the algorithm is not penalized when the ordering of papers induced by the similarity scores $\{\similarity_{\participant}^{(i)}, \similarity_{\participant}^{(j)}\}$ agrees with the ground truth expertise-based ordering $\{\expertise_{\participant}^{(i)}, \expertise_{\participant}^{(j)}\}$. When two orderings disagree (that is, the algorithm makes an error), the algorithm is penalized by the difference of expertise reported by the participant $( |\expertise_{\participant}^{(i)} - \expertise_{\participant}^{(j)} | )$. Finally, when scores computed by the algorithm indicate a tie while expertise scores are different, the algorithm is penalized by half the difference in expertise $(\frac{1}{2} |\expertise_{\participant}^{(i)} - \expertise_{\participant}^{(j)} | )$.

Having the unnormalized loss with respect to a participant defined, we now compute the overall loss $\loss \in [0, 1]$ 
of the algorithm. For this, we take the sum of unnormalized losses across all participants and normalize this sum by the loss achieved by the adversarial algorithm that reverses the ground-truth ordering of expertise (that is, sets $\similarity = -\expertise$) and achieves the worst possible performance on the task. More formally,

\begin{align*}
    \loss = \frac{\sum\limits_{\participant} \loss_\participant}{\sum\limits_{\participant} \sum\limits_{\substack{i, j = 1 \\ i < j}}^{\numpapers_\participant} \big |\expertise_{\participant}^{(i)} - \expertise_{\participant}^{(j)} \big |}.
\end{align*}
Overall, our loss $\loss$ takes values from $0$ to $1$ with lower values indicating better performance. 
\subsection{Algorithms}
\label{section:setup:algorithms}

In this work, we evaluate several algorithms that we now discuss. All of these algorithms operate with (i) the list of submissions for which similarity scores need to be computed and (ii) reviewers' profiles comprising past publications of reviewers. Conceptually, all algorithms predict reviewers' expertise by evaluating textual overlap between each submission and papers in each reviewer's publication profile. Let us now provide more detail on how this idea is implemented in each of the algorithms under consideration.

\paragraph{Trivial baseline} First, we consider a trivial baseline that ignores the content of submissions and reviewers' profiles when computing the assignment scores: for each (participant, paper) pair, the \constant{} algorithm predicts the score $\similarity$ to be 1.

\paragraph{Toronto Paper Matching System (TPMS)} The \tpms{} algorithm~\citep{charlin13tpms}, which is based on TF-IDF similarities, is widely used by flagship conferences such as ICML and AAAI. While an exact implementation is not publicly available, in our experiments we use an open-source version by~\citet{xu2018strategyproof} which implements the basic TF-IDF logic of \tpms{}. We also note that the TF-IDF method was also independently proposed earlier by~\citet{price2010subsift}, and subsequently also rediscovered by~\citet{kerzendorf2020distributed}. 

As a technical remark, in our implementation we use reviewers' profiles and all reported papers to compute the IDF part of the TF-IDF model. In principle, one may be able to get a better performance by using a larger set of papers from the respective field (e.g., all submissions to the last edition of the conference) to compute IDF.

\paragraph{OpenReview algorithms} OpenReview is a conference-management system used by machine learning conferences such as NeurIPS and ICLR. It offers a family of deep-learning algorithms for measuring expertise of reviewers in reviewing submissions. In this work, we evaluate the following algorithms which are used to compute affinity scores between submissions and reviewers:

\begin{itemize}
    \item \elmo{}. This algorithm relies on general-purpose \textbf{E}mbeddings from \textbf{L}anguage \textbf{Mo}dels~\citep{peters18deep} to compute textual similarities between submissions and reviewers' past papers.
    
    \item \specter{}. This algorithm employs more specialized document-level embeddings of scientific documents~\citep{cohan2020specter}. Specifically, \specter{} explores the citation graph to construct embeddings that are useful for a variety downstream tasks, including the similarity computation we focus on in this work.
    
    \item \mfr{}. \mfr{}  enhances \specter{}~\citep{chang2021cold}. Instead of constructing a single embedding of each paper, it constructs multiple embeddings that correspond to different facets of the paper. These embeddings are then used to compute the similarity scores.

    \item \spectertwo{}. Finally, \spectertwo{} ~\citep{Singh2022SciRepEvalAM} is an updated version that builds on top of  \specter{}~\citep{chang2021cold}. They employ the use of adapters for the embeddings to be more specialized for specific downstream scientific tasks.
\end{itemize}

We use implementations of these methods that are available on the OpenReview GitHub page\footnote{\url{https://github.com/openreview/openreview-expertise}} and execute them with default parameters.

When implementing these algorithms, an important design decision is how to aggregate—or ``pool''—the similarity scores between a reviewer's profile and a submitted paper. In the openreview toolkit, this pooling is done by first computing the similarity score between the submitted paper and each individual paper in the reviewer's profile. These scores are then aggregated into a single similarity measure using a pooling method. Common pooling methods include taking the maximum or the mean of the individual scores. An alternative, suggested by~\citet{Hsieh_Raghunathan_Shah_2024}, involves using a percentile—such as the 75th percentile—to improve robustness (as compared to max) against fraudulent behavior.  We evaluate all three pooling methods in the context of \spectertwo{}.
 
\paragraph{ACL paper matching} The Association for Computational Linguistics (ACL) is a community of researchers working on computational problems involving natural language. The association runs multiple publication venues, including the flagship ACL conference, and has its own method to compute expertise between papers and reviewers.\footnote{\url{https://github.com/acl-org/reviewer-paper-matching}} 
This algorithm is trained on 45,309 abstracts of papers found in the ACL anthology (\url{https://aclanthology.org}) downloaded in November 2019. The key idea of the model training is to split each abstract into two non-contiguous parts and treat two parts of the same abstract as a positive example and highly similar parts of different abstracts as negative examples. The algorithm uses ideas from the work of~\citet{wieting-etal-2019-simple,wieting-etal-2022-paraphrastic} to learn a simple and efficient similarity function that separates positive examples from negative examples. This function is eventually used to compute similarity scores between papers and reviewers. More details on the \acl{} algorithm are provided in Appendix~\ref{section:acl}.

We note that the \acl{} algorithm is trained on the domain of computational linguistics and hence may suffer from a distribution shift when applied to the general machine learning domain. That said, in line with \elmo{}, \specter{}, and \mfr{}, we use the \acl{} algorithm without additional fine-tuning. 

\paragraph{Off-the-Shelf LLMs} Large Language Models (LLMs) have increasingly shown high performance on specific downstream tasks. To this end, we evaluate commonly used LLMs---\openai{}, \claude{}, and \gemini{}---on this dataset.\footnote{The servers for DeepSeek were consistently overloaded, and Grok3 APIs were unavailable.} We query the models with a prompt that asks the model to assign a score to any given reviewer-paper pair. The specific details are outlined in Appendix~\ref{section:prompt}.


\section{Results}

\label{section:results}

In this section, we report the results of evaluation of algorithms described in Section~\ref{section:setup:algorithms}. First, we juxtapose all algorithms on our data (Section~\ref{section:results:comparison}). Second, we use the \tpms{} algorithm to explore various aspects of the similarity-computation problem (Section~\ref{section:results:exploration}).

Before presenting the results, we note that in this section, we conduct all evaluations focusing on papers that have PDFs freely available. To this end, we remove 6 papers from the dataset as they are not freely available online (see Table~\ref{table:papers_demo}). Similarly, we limit reviewer profiles to papers whose semantic scholar profiles contain links to arXiv. One of the participants did not have any such papers, so we exclude them from the dataset.

\subsection{Comparison of the algorithms}
\label{section:results:comparison}

Our first set of results compares the performance of the existing similarity-computation algorithms. To run these algorithms on our data, we need to make some modeling choices faced by conference organizers in practice:
\begin{itemize}[leftmargin=20pt, itemsep=0pt, topsep=5pt]
    \item \textit{Paper representation.} First, in their inner-workings, similarity-computation algorithms operate with some representation of the paper content. Possible choices of representations include: (i) title of the paper, (ii) title and abstract, and (iii) full text of the paper. We choose option (ii) as this option is often used in real conferences and is supported by all algorithms we consider in this work. Thus, to predict expertise, algorithms are provided with the title and abstract of each paper (both papers submitted to the conference and papers in reviewers' publication profiles).
    
    \item \textit{Reviewer profiles.} The second important parameter is the choice of papers to include in reviewers' profiles. In real conferences, this choice is often left to reviewers who can manually select the papers they find representative of their expertise. In our experiments, we construct reviewer profiles automatically by using the 20 most recent papers from their Semantic Scholar profiles. If a reviewer has less than 20 papers published, we include all of them in their profile. Our choice of the reviewer profile size is governed by the observation that the mean length of the reviewer profile in the NeurIPS 2022 conference is 16.5 papers. By setting the maximum number of papers to 20, we achieve the mean profile length of 14.8, thereby operating with the amount of information close to that available to algorithms in real conferences. 
\end{itemize}

\paragraph{Statistical aspects} To build reviewer profiles, we use publication years to order papers by recency, where we break ties uniformly at random. Thus, the content of reviewer profiles depends on randomness. To average this randomness out, we repeat the procedure of profile construction and similarity prediction 10 times, and report the mean loss over these iterations. That said, we note that the observed variability due to the randomness in the construction of reviewer profiles is negligible, with a standard deviation over all iterations is less than 0.005.

The pointwise performance estimates obtained by the procedure above depend on the selection of participants who contributed to our dataset. To quantify the associated level of uncertainty, we compute 95\% confidence intervals as follows. For 1,000 iterations, we create a new \emph{reviewer pool} by sampling participants with replacement and recomputing the loss of each algorithm on the bootstrapped set of reviewers. To save computation time, we do not reconstruct reviewer profiles for each of these iterations as the uncertainty associated with the construction of reviewer profiles is small. Instead, we reuse profiles constructed to obtain pointwise estimates.

Finally, we also build confidence intervals for the \emph{difference} in the performance of the algorithms. We do so in addition to the aforementioned procedure since even when the losses of the algorithms fluctuate with the choice of the bootstrapped dataset, the relative difference in performance of a pair of algorithms may be stable. Specifically, we use the procedure above to build confidence intervals for the difference in performance between the \tpms{} algorithm and each of the more advanced algorithms. \tpms{} is chosen as a baseline for this comparison due to its simplicity.

\medskip

\paragraph{Results of the comparison} Table~\ref{table:results_main} displays results of the comparison. The first pair of columns presents the loss of each algorithm on our dataset and the associated confidence intervals. The third and the forth columns investigate the relative difference in performance between the non-trivial algorithms: they display the differences in performance between \tpms{} and each of the more advanced algorithms (OpenReview algorithms and \acl{}) together with the associated confidence intervals. We make several important observations from Table~\ref{table:results_main}:

\begin{table*}[t]
\begin{center}
\begin{small}
\begin{sc}
\begin{tabular}{lcccc}
\toprule
Algorithm &  {Loss} & {95\% CI for Loss} & {$\Delta$ with \tpms{}} & { 95\% CI for $\Delta$} \\
\midrule
\constant{} & $0.50$ & --- & --- & ---\\ 
\midrule
\tpms{} & $0.28$ & $[0.23, 0.33]$ & --- & --- \\
\midrule
\elmo{} & $0.35$ & $[0.31, 0.42]$ & $+0.07$ & $[0.02, 0.14]$ \\
\specter{} & $0.27$ & $[0.21, 0.34]$ & $-0.01$ & $[-0.06, 0.04]$ \\
\mfr{} & $0.24$ & $[0.18, 0.30]$ & $-0.04$ & $[-0.09, 0.01]$ \\
\spectertwo{} Mean Pool& $0.25$ & $[0.19, 0.30]$ & $-0.03$ & $[-0.1, 0.02]$ \\
\spectertwo{} 75th Pool& $0.24$ & $[0.19, 0.30]$ & $-0.04$ & $[-0.1, 0.02]$ \\
\spectertwo{} Max Pool& \textbf{0.22} & $[0.18, 0.26]$ & $-0.06$ & $[-0.11, -0.02]$ \\
\midrule
\acl{} & $0.30$ & $[0.25, 0.36]$ & $+0.02$ & $[-0.02, 0.07]$\\
\midrule
\openai{} & $0.25$ & $[0.21, 0.29]$ & $-0.03$ & $[-0.07, 0.0]$ \\
\claude{} & $0.31$ & $[0.27, 0.35]$  & $+0.03$ & $[-0.01, 0.07]$ \\
\gemini{} & $0.38$ & $[0.34, 0.41]$ & $+0.10$ & $[0.06, 0.14]$ \\
\bottomrule
\end{tabular}
\end{sc}
\end{small}
\end{center}
\caption{Comparison of similarity-computation algorithms on the collected data. All algorithms operate with reviewer profiles consisting of the 20 most recent papers and use titles and abstracts of papers. Lower values of loss are better. A positive (respectively, negative) value of $\Delta$ indicates that the algorithm performs worse (respectively, better) than \tpms{}.${}^{*}$\\ \begin{footnotesize}
${}^{\ast}$Later in Table~\ref{table:results_explore} we evaluate \tpms{} algorithm when full texts of papers are additionally provided and observe that \tpms{} closes the gap with \spectertwo{} when using this information. 
\end{footnotesize}
}
\label{table:results_main}
\end{table*}

\begin{itemize}[leftmargin=20pt, itemsep=0pt, topsep=5pt]
    \item First, we note that all algorithms we consider in this work considerably outperform the \constant{} baseline, confirming that content of papers is useful in evaluating expertise. 
    
    \item Second, comparing the four algorithms from the OpenReview toolkit, we note that the \specter{} family of algorithms outperform \elmo{}. The former rely on domain-specific embeddings\footnote{\mfr{}, \specter{}, and \spectertwo{} are initialized with SciBERT~\citep{beltagy-etal-2019-scibert}} while \elmo{} uses general-purpose embeddings. Thus, the nature of the text similarity computation task in the academic context may be sufficiently different from that in other domains.

    \item Third, we note that under modeling choices (paper representation and length of reviewers' profiles) that mimic choices that have been made in real conferences, the \spectertwo{} algorithms performs the best, with \spectertwo{} with max pooling achieving the lowest loss.

    \item Fourth, we note that generic LLMs have a large variation of performance. While \openai{} performs on par with the \spectertwo{} algorithm, \gemini{} is the worst performing algorithm barring the \constant{} baseline.

    \item Fifth, among the pooling options in \spectertwo{}, max pooling delivers the best performance, while the 75th percentile pooling method performs nearly as well.
    
   \item The sixth finding is the most surprising. Observe that the \tpms{} algorithm is much simpler than other non-trivial algorithms: in contrast to \elmo{}, \specter{}, \acl{}, and the LLMs, it does not rely on carefully learned embeddings. However, \tpms{} is competitive against complex \spectertwo{} and \mfr{} and even outperforms \claude{} and \acl{}. Moreover, \tpms{} is the only algorithm in the OpenReview toolkit that can make use of the full text of the papers. In Section~\ref{section:results:exploration}, we find that when \tpms{} uses the full text of all papers (including reviewers' past papers), it closes the gap with \spectertwo{}.
\end{itemize}


Finally, we note that some of the confidence intervals for the performance of the different algorithms, as well as for the relative differences, are overlapping. It is, therefore, crucial to increase the size of our dataset to enable more fine-grained comparisons between the algorithms.

\subsection{The role of modeling choices}
\label{section:results:exploration}

In the beginning of Section~\ref{section:results:comparison} we made two modeling choices pertaining to (i) representations of the papers provided to similarity-computation algorithms, and (ii) the size of reviewers' profiles used by these algorithms. In this section, we investigate these two questions in more detail.

\begin{itemize}[leftmargin=20pt, itemsep=0pt, topsep=5pt]
    \item \textit{Question 1 (paper representation).} Some similarity-computation algorithms are designed to work with titles and/or abstracts of papers (e.g., \specter) while others can also incorporate the full texts of the manuscripts (e.g., \tpms{}). Consequently, there is a potential trade-off between accuracy and computation time. Richer representations are envisaged to result in higher accuracy, but are also associated with an increased demand for computational power. As with the choice of the algorithm itself, there is no guidance on what amount of information should be supplied to the algorithm as the gains from using more information are not quantified. With this motivation, our first question is:
    \begin{quote}
        \emph{What are the benefits of providing richer representations of papers to similarity-computation algorithms?}
    \end{quote}
    
    \item \textit{Question 2 (reviewer profile).} The second important choice is the size of reviewers' profiles. On the one hand, by including only very recent papers in reviewers' profiles, conference organizers are at risk of not using enough data to obtain accurate values of expertise. On the other hand, old papers may not accurately represent the current expertise of researchers and hence may result in noise when used to compute expertise. Thus, our second question is:

    \begin{quote}
        \emph{What is the optimal number of the most recent papers to include in the profiles of reviewers?}
    \end{quote}

\end{itemize}

To investigate these questions, we choose the \tpms{} algorithm as the workhorse to perform experiments. We make this choice for two reasons. First, \tpms{} can work with all possible choices of the paper representation: title only, title and abstract, and full text of the paper. In contrast, other methods do not support using the full text of the papers. Second, \tpms{} is fast to execute, enabling us to compute expertise for hundreds of  configurations in a reasonable time.

With the algorithm chosen, we vary the number of papers in the reviewers' profiles from 1 to 20. For each value of the profile length, we consider three representations of the paper content: title, title+abstract, and full text of the paper. For each combination of parameters, we construct reviewer profiles and predict similarities using the approach introduced in Section~\ref{section:results:comparison}. The only exception is that we repeat the procedure for averaging out the randomness in the profile creation 5 times (instead of 10) to save the computation time.

\begin{figure}[tb]
    \centering
    \includegraphics[width=10cm]{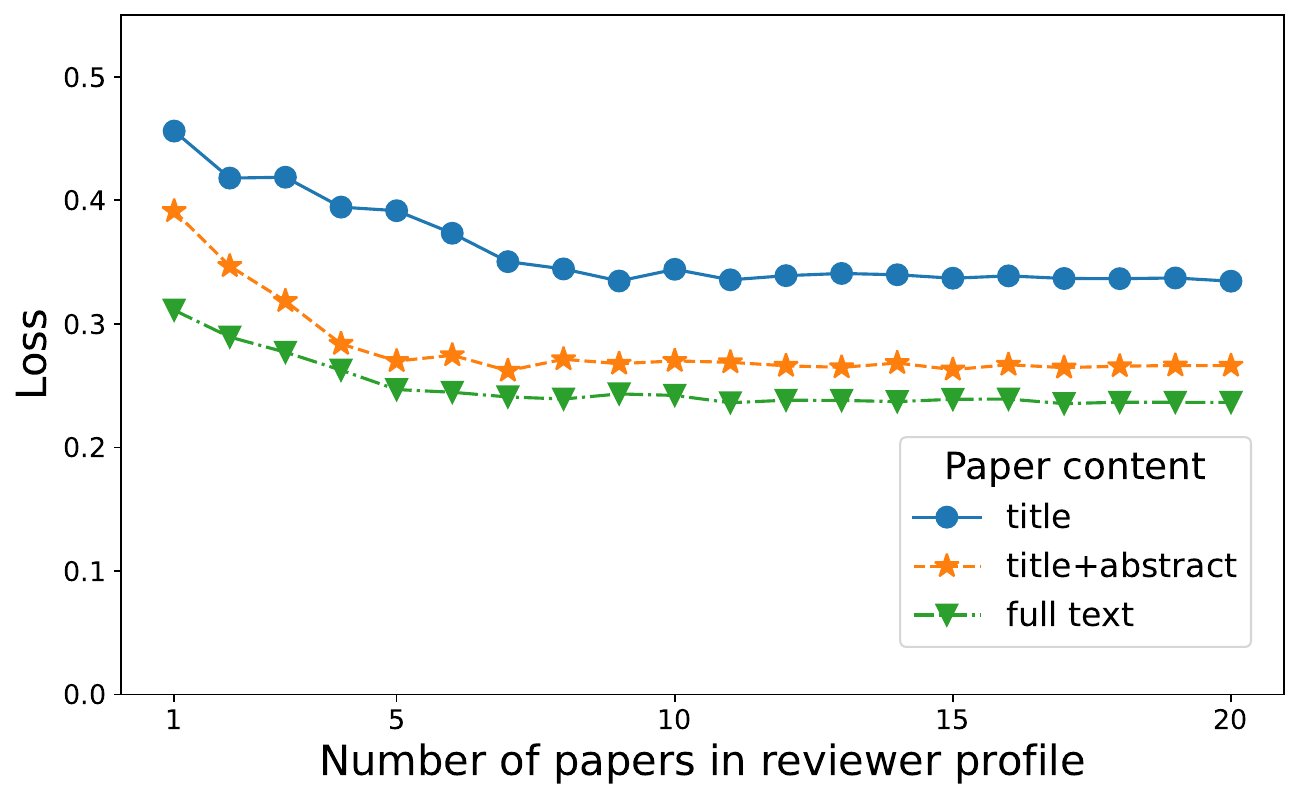}
    \caption{Impact of different choices of parameters on the quality of predicted similarities. Confidence intervals are not shown (see Table~\ref{table:results_explore}).}
    \label{fig:results_explore}
\end{figure}

\paragraph{Results} Figure~\ref{fig:results_explore} and Table~\ref{table:results_explore} provide answers to the questions we study in this section. Figure~\ref{fig:results_explore} shows the pointwise loss of the \tpms{} algorithm for each choice of parameters. To save computation time, we do not build confidence intervals for each combination of parameters. Instead, Table~\ref{table:results_explore} sets the number of papers in reviewers' profiles to 20 (consistent with Table~\ref{table:results_main}) and presents confidence intervals for losses incurred by the algorithm under different choices of paper representations. We now make two observations.

\begin{table*}[tb]
\begin{center}
\begin{small}
\begin{sc}
\begin{tabular}{lcccc}
\toprule
 Paper Representation &  Loss & 95\% CI for Loss & $\Delta$ with Title+Abstract & 95\%
 CI for $\Delta$ \\
  \midrule
Title & $0.33$ & $[0.29, 0.38]$ & $+0.07$ & $[0.02, 0.12]$ \\
Title+Abstract & $0.27$ & $[0.22, 0.32]$ & --- & --- \\
Full Text & \textbf{0.24}  & $[0.19, 0.30]$ & $-0.03$ & $[-0.09, 0.03]$ \\
\bottomrule
\end{tabular}
\end{sc}
\end{small}
\end{center}
\caption{Performance of the \tpms{} algorithm with 20 most recent papers included in reviewers' profiles and with different choices of the paper representation. Lower values of loss are better. A positive (respectively, negative) value of $\Delta$ indicates that the corresponding choice of the paper representation leads to a weaker (respectively, stronger) performance.}
\label{table:results_explore}
\end{table*}

First, paper abstracts are very useful in improving the quality of expertise prediction as compared to titles alone (the improvement of $0.07$). Including full texts of the papers in reviewers' and papers' profiles results in additional improvement ($0.03$). Overall, ignoring the minor differences in the datasets between this section and Section~\ref{section:results:comparison}, we observe that \tpms{} with titles, abstracts, and full texts of papers demonstrates a performance quite close to \spectertwo{} which cannot handle the full texts of papers. Thus, it may be of interest to develop modern deep-learning based algorithms with longer contexts to incorporate full texts of papers, and further boost performance on the similarity-computation task.

Second, the loss curves plateau once reviewer profiles include 8 or more of their most recent papers. Additional increase of the profile length does not impact the quality of predictions. Thus, in practice, reviewers may be instructed to include 10 representative papers to their profile, which for many active researchers amounts to the number of papers published in 1-3 years.

\section{Easy versus hard stratification}
\label{section:section:easyhard}

Recall that our loss metric (see Section~\ref{section:expsetup}) assigns different weights to errors of an algorithm: errors on pairs with similar values of reviewer's expertise are penalized less than errors on well-separated pairs. The overall performance of the algorithm is then captured in a single number (loss) which does not characterize the type of mistakes made by the algorithm.

To provide deeper characterization of the algorithm performance, we now conduct additional evaluations. In these evaluations, we focus on two important regions of the similarity-computation problem. Specifically, from all expertise evaluations made by participants, we select two groups of triplets, where each triplet consists of a participant $\participant$ and two papers $(\paper_1, \paper_2)$ evaluated by this participant:
\begin{itemize}
    \item \textbf{``Easy'' triplets.} The first group comprises 261 triplets where a participant reported high expertise (greater than or equal to four) in one paper and low expertise (less than or equal to two) in another.

    We call these triplets ``easy'' as the gap in the expertise between the two papers is large. Note that in real conferences it is important to resolve the easy triplets correctly to ensure that reviewers do not get assigned irrelevant papers.

    \item \textbf{``Hard'' triplets.} The second group comprises 417 triplets where a participant reported high expertise (greater than or equal to four) in both papers and the values of expertise are different for the two papers.

    We call these triplets ``hard'' as the gap in the expertise between the papers is small. That said, we note that resolving hard triplets in practice is also very important: in real conferences, we want to assign papers to the most suitable reviewers and we need to be able to distinguish two papers with high but not equal values of expertise to achieve this goal.
\end{itemize}

Note that by focusing on these two groups, we exclude regions of the problem which may be considered less important. For example, the ability of an algorithm to correctly order two low-expertise papers is less crucial as long as the algorithm can reliably distinguish these papers from the high-expertise papers. 

We now study the performance of the algorithms introduced in Section~\ref{section:setup:algorithms} on the easy and hard groups of triplets. For all algorithms we use titles and abstracts of papers as the data source and include 20 papers in the reviewers' profiles. Specifically, for each triplet we compare the ordering of papers predicted by an algorithm with the expertise-induced ordering. We then compute the accuracy of each algorithm as the fraction of triplets correctly resolved by the algorithm (from 0 to 1, larger values are better). In addition, we also evaluate \tpms{} in the full text regime to estimate the effect of papers' texts on the accuracy of the similarity-computation algorithms.\footnote{\tpms{} in the full text regime is evaluated on a slightly different dataset as explained in Section~\ref{section:results:exploration}.}

Table~\ref{table:additional_eval} demonstrates the results of additional evaluations. First, note that all algorithms have moderate to high accuracy in resolving the easy triplets. Most algorithms are able to detect papers that reviewers have low expertise in with probability close to or above 80\%, with \spectertwo{} reaching nearly 90\% accuracy. Surprisingly, LLMs \gemini{} and \claude{}, along with \elmo{} are the 3 with the lowest performance, with \gemini{} having as low as 57\% accuracy. However, we note that in a non-trivial amount of cases (more than 10\%), the algorithms are unable to distinguish a paper that a reviewer is well-qualified to review versus a paper the reviewer does not have necessary background to evaluate. Thus, there is a vital need to improve the similarity-computation algorithms.

Getting to the hard triplets, we observe a significant drop in performance with most algorithms being a bit better than random: the best-performing algorithm (\tpms{}) reaches 62\% accuracy in the title+abstract regime and 64\% in the full text regime. Perhaps surprisingly, while the \specter{} family of algorithms outperform \tpms{} on easy triplets, they fare slightly worse than \tpms{} on hard triplets (we caveat though that error bars are too wide to make a decisive comparison). Furthermore, the performance of all three LLMs are no better than random, with \gemini{} performing worse than random. This observation suggests that there may be a value in additionally fine-tuning these advanced algorithms on hard triplets to improve their practical performance.

Finally, we observe that on both easy and hard triplets, full texts of the papers are instrumental in improving performance of the \tpms{} algorithm. This observation supports our intuition that similarity-computation algorithms do indeed benefit from employing full texts of papers.

\medskip

\noindent With these remarks, we conclude evaluations of algorithms on the dataset we collected.

\begin{table*}[tb]
\begin{center}
\begin{small}
\begin{sc}
\begin{tabular}{lcccc}
\toprule
& \multicolumn{2}{c}{Easy triplets ($n  = 261$)} & \multicolumn{2}{c}{Hard triplets ($n  = 417$)} \\
\cmidrule(r){2-3} 
\cmidrule(r){4-5}
Algorithm &  {Accuracy} & {95\% CI} & {Accuracy} & {95\% CI} \\
\midrule
\tpms{} (T+A) & $0.80$ & $[0.72, 0.87]$ & $0.62$ & $[0.54, 0.69]$ \\
\tpms{} (Full Text) & $0.84$ & $[0.76, 0.91]$ & \textbf{0.64} & $[0.56, 0.70]$ \\
\midrule
\elmo{} & $0.70$ & $[0.62, 0.78]$ & $0.57$ & $[0.51, 0.63]$ \\
\specter{} & $0.85$ & $[0.76, 0.92]$ & $0.57$ & $[0.50, 0.63]$ \\
\mfr{} & $0.88$ & $[0.81, 0.94]$ & $0.60$ & $[0.53; 0.66]$ \\
\spectertwo{} Mean Pool & $0.84$ & $[0.73, 0.92]$ & $0.59$ & $[0.53, 0.65]$ \\
\spectertwo{} 75th Pool & $0.85$ & $[0.75, 0.93]$ & $0.60$ & $[0.54, 0.65]$ \\
\spectertwo{} Max Pool & $\textbf{0.89}$ & $[0.82, 0.94]$ & $0.61$ & $[0.54, 0.67]$ \\
\midrule
\acl{} & $0.78$ & $[0.69; 0.86]$ & $0.62$ & $[0.55; 0.68]$\\
\midrule
\openai{} & $0.82$ & $[0.75, 0.88]$ & $0.52$ & $[0.47, 0.57]$ \\
\claude{} & $0.72$ & $[0.66, 0.77]$  & $0.52$ & $[0.47, 0.58]$ \\
\gemini{} & $0.57$ & $[0.50, 0.63]$ & $0.41$ & $[0.38, 0.43]$ \\
\bottomrule
\end{tabular}
\end{sc}
\end{small}
\end{center}
\caption{Results of additional evaluations. Higher values of accuracy are better. \spectertwo{} demonstrates the best performance on easy triplets and \tpms{} (Full Text) has the highest accuracy on hard triplets.}
\label{table:additional_eval}
\end{table*}

\section{Discussion}
\label{section:discussion}

In this work, we collect a novel dataset of reviewers' expertise in reviewing papers. In contrast to datasets collected in previous works, our dataset (i) is released publicly, and (ii) contains  evaluations of expertise made by scientists who have actually read the papers for their own research purposes. We use this dataset to compare several existing expertise-estimation algorithms and help venue organizers in choosing an algorithm in an evidence-based manner. 

Most importantly, we find that current algorithms make a large number of errors. Given the growing number of submissions in many fields of research, the need for automation in reviewer assignments is only growing. There is thus an urgent and vital need to develop significantly improved algorithms to match reviewers to papers, thereby in turn making the peer-review process considerably better.

Our dataset can be used to develop as well as evaluate new expertise-estimation algorithms. We encourage researchers from the natural language processing and other communities to use our data in order to improve peer review.

\paragraph{Limitations.} Finally, we mention several caveats that researchers should be aware of when working with our dataset and interpreting the results of our experiments. First, our dataset comprises evaluations of expertise in reviewing papers that were written some time ago. In contrast, in real conferences, many papers are recent and not available online. Thus, incoming citations to papers included in our dataset may constitute information that is not available to the algorithms in real life. While the algorithms we evaluate in this work do not rely on the citation relationship, this caveat may be important for future work.

Second, the experiments we conduct in this work rely on Semantic Scholar profiles. These profiles are constructed automatically and may not be accurate. Thus, mistakes of the algorithms we observe in this work can be partially due to the noise in the profile creation.

Third, the design of the survey interface can influence the data and, consequently, the evaluations. We hope for further studies on reviewer expertise, potentially employing diverse methodologies. When combined with our dataset, these additional studies could provide a more robust and multi-dimensional evaluation.

Finally, we reiterate that the present version of the dataset can have sampling bias and may be constructed by participants who collectively are not representative of the general computer science community. For example, about 40\% of participants are affiliated with \cmu. To alleviate this issue, we continue the data collection process and encourage the readers of this paper to contribute their data points to the dataset:
\begin{center}
    \weblink
\end{center}

\section*{Acknowledgments}
We sincerely thank all participants of our survey for their contribution to the dataset. This work was supported by NSF CAREER 1942124 and NSF 1763734.

\bibliographystyle{tmlr}

\begin{thebibliography}{50}
	\providecommand{\natexlab}[1]{#1}
	\providecommand{\url}[1]{\texttt{#1}}
	\expandafter\ifx\csname urlstyle\endcsname\relax
	\providecommand{\doi}[1]{doi: #1}\else
	\providecommand{\doi}{doi: \begingroup \urlstyle{rm}\Url}\fi
	
	\bibitem[Anjum et~al.(2019)Anjum, Gong, Bhat, Hwu, and Xiong]{anjum19pare}
	Omer Anjum, Hongyu Gong, Suma Bhat, Wen-Mei Hwu, and JinJun Xiong.
	\newblock {P}a{R}e: A paper-reviewer matching approach using a common topic
	space.
	\newblock In \emph{Proceedings of the 2019 Conference on Empirical Methods in
		Natural Language Processing and the 9th International Joint Conference on
		Natural Language Processing (EMNLP-IJCNLP)}, pp.\  518--528, Hong Kong,
	China, November 2019. Association for Computational Linguistics.
	\newblock \doi{10.18653/v1/D19-1049}.
	\newblock URL \url{https://aclanthology.org/D19-1049}.
	
	\bibitem[Beltagy et~al.(2019)Beltagy, Lo, and Cohan]{beltagy-etal-2019-scibert}
	Iz~Beltagy, Kyle Lo, and Arman Cohan.
	\newblock {S}ci{BERT}: A pretrained language model for scientific text.
	\newblock In \emph{Proceedings of the 2019 Conference on Empirical Methods in
		Natural Language Processing and the 9th International Joint Conference on
		Natural Language Processing (EMNLP-IJCNLP)}, pp.\  3615--3620, Hong Kong,
	China, November 2019. Association for Computational Linguistics.
	\newblock \doi{10.18653/v1/D19-1371}.
	\newblock URL \url{https://aclanthology.org/D19-1371}.
	
	\bibitem[Bianchi \& Squazzoni(2015)Bianchi and Squazzoni]{bianchi2015three}
	Federico Bianchi and Flaminio Squazzoni.
	\newblock Is three better than one? {S}imulating the effect of reviewer
	selection and behavior on the quality and efficiency of peer review.
	\newblock In \emph{Proceedings of the 2015 Winter Simulation Conference}, pp.\
	4081--4089. IEEE Press, 2015.
	
	\bibitem[Black et~al.(1998)Black, Van~Rooyen, Godlee, Smith, and
	Evans]{black1998makes}
	Nick Black, Susan Van~Rooyen, Fiona Godlee, Richard Smith, and Stephen Evans.
	\newblock What makes a good reviewer and a good review for a general medical
	journal?
	\newblock \emph{Jama}, 280\penalty0 (3):\penalty0 231--233, 1998.
	
	\bibitem[Boehmer et~al.(2021)Boehmer, Bredereck, and
	Nichterlein]{boehmer2021combating}
	Niclas Boehmer, Robert Bredereck, and Andr{\'e} Nichterlein.
	\newblock Combating collusion rings is hard but possible.
	\newblock \emph{arXiv preprint arXiv:2112.08444}, 2021.
	
	\bibitem[Chang \& McCallum(2021)Chang and McCallum]{chang2021cold}
	Haw-Shiuan Chang and Andrew McCallum.
	\newblock Cold-start paper recommendation using multi-facet embedding.
	\newblock \emph{Overleaf preprint}, 2021.
	\newblock \url{https://www.overleaf.com/project/5f359923225f06000134ea95} [Last
	Accessed: 3/15/2023].
	
	\bibitem[Charlin \& Zemel(2013)Charlin and Zemel]{charlin13tpms}
	L.~Charlin and R.~S. Zemel.
	\newblock The {T}oronto {P}aper {M}atching {S}ystem: An automated
	paper-reviewer assignment system, 2013.
	\newblock URL \url{http://www.cs.toronto.edu/~zemel/documents/tpms.pdf}.
	
	\bibitem[Cohan et~al.(2020)Cohan, Feldman, Beltagy, Downey, and
	Weld]{cohan2020specter}
	Arman Cohan, Sergey Feldman, Iz~Beltagy, Doug Downey, and Daniel~S. Weld.
	\newblock Specter: Document-level representation learning using
	citation-informed transformers, 2020.
	
	\bibitem[Dhull et~al.(2022)Dhull, Jecmen, Kothari, and Shah]{dhull2022price}
	Komal Dhull, Steven Jecmen, Pravesh Kothari, and Nihar~B Shah.
	\newblock Strategyproofing peer assessment via partitioning: The price in terms
	of evaluators’ expertise.
	\newblock In \emph{HCOMP}, 2022.
	
	\bibitem[Dumais \& Nielsen(1992)Dumais and Nielsen]{dumais1992automating}
	Susan~T Dumais and Jakob Nielsen.
	\newblock Automating the assignment of submitted manuscripts to reviewers.
	\newblock In \emph{Proceedings of the 15th annual international ACM SIGIR
		conference on Research and development in information retrieval}, pp.\
	233--244, 1992.
	
	\bibitem[Fiez et~al.(2020)Fiez, Shah, and Ratliff]{fiez2019super}
	T~Fiez, N~Shah, and L~Ratliff.
	\newblock A {SUPER}* algorithm to optimize paper bidding in peer review.
	\newblock In \emph{Conference on Uncertainty in Artificial Intelligence}, 2020.
	
	\bibitem[Garg et~al.(2010)Garg, Kavitha, Kumar, Mehlhorn, and
	Mestre]{Garg2010papers}
	N.~Garg, T.~Kavitha, A.~Kumar, K.~Mehlhorn, and J.~Mestre.
	\newblock Assigning papers to referees.
	\newblock \emph{Algorithmica}, 58\penalty0 (1):\penalty0 119--136, Sep 2010.
	\newblock ISSN 1432-0541.
	\newblock \doi{10.1007/s00453-009-9386-0}.
	\newblock URL \url{https://doi.org/10.1007/s00453-009-9386-0}.
	
	\bibitem[Goldsmith \& Sloan(2007)Goldsmith and Sloan]{goldsmith07aiconf}
	Judy Goldsmith and {Robert H.} Sloan.
	\newblock The {AI} conference paper assignment problem.
	\newblock \emph{AAAI Workshop - Technical Report}, WS-07-10:\penalty0 53--57,
	12 2007.
	
	\bibitem[Jecmen et~al.(2020)Jecmen, Zhang, Liu, Shah, Conitzer, and
	Fang]{jecmen2020manipulation}
	Steven Jecmen, Hanrui Zhang, Ryan Liu, Nihar~B. Shah, Vincent Conitzer, and Fei
	Fang.
	\newblock Mitigating manipulation in peer review via randomized reviewer
	assignments.
	\newblock In \emph{NeurIPS}, 2020.
	
	\bibitem[Jecmen et~al.(2022)Jecmen, Yoon, Conitzer, Shah, and
	Fang]{jecmen22dataset}
	Steven Jecmen, Minji Yoon, Vincent Conitzer, Nihar~B. Shah, and Fei Fang.
	\newblock A dataset on malicious paper bidding in peer review, 2022.
	\newblock URL \url{https://arxiv.org/abs/2207.02303}.
	
	\bibitem[Jhih-Yi et~al.(2024)Jhih-Yi, Hsieh, Raghunathan, and
	Shah]{Hsieh_Raghunathan_Shah_2024}
	Jhih-Yi, Hsieh, Aditi Raghunathan, and Nihar~B. Shah.
	\newblock Vulnerability of text-matching in ml/ai conference reviewer
	assignments to collusions.
	\newblock \penalty0 (arXiv:2412.06606), December 2024.
	\newblock \doi{10.48550/arXiv.2412.06606}.
	\newblock URL \url{http://arxiv.org/abs/2412.06606}.
	\newblock arXiv:2412.06606.
	
	\bibitem[Kendall(1938)]{kendall1938new}
	Maurice~G Kendall.
	\newblock A new measure of rank correlation.
	\newblock \emph{Biometrika}, 30\penalty0 (1/2):\penalty0 81--93, 1938.
	
	\bibitem[Kerzendorf et~al.(2020)Kerzendorf, Patat, Bordelon, van~de Ven, and
	Pritchard]{kerzendorf2020distributed}
	Wolfgang~E. Kerzendorf, Ferdinando Patat, Dominic Bordelon, Glenn van~de Ven,
	and Tyler~A. Pritchard.
	\newblock Distributed peer review enhanced with natural language processing and
	machine learning, 2020.
	
	\bibitem[Kobren et~al.(2019)Kobren, Saha, and McCallum]{kobren19localfairness}
	Ari Kobren, Barna Saha, and Andrew McCallum.
	\newblock Paper matching with local fairness constraints.
	\newblock In \emph{ACM SIGKDD International Conference on Knowledge Discovery
		\& Data Mining}, 2019.
	
	\bibitem[Kudo \& Richardson(2018)Kudo and
	Richardson]{kudo-richardson-2018-sentencepiece}
	Taku Kudo and John Richardson.
	\newblock {S}entence{P}iece: A simple and language independent subword
	tokenizer and detokenizer for neural text processing.
	\newblock In \emph{Proceedings of the 2018 Conference on Empirical Methods in
		Natural Language Processing: System Demonstrations}, pp.\  66--71, Brussels,
	Belgium, November 2018. Association for Computational Linguistics.
	\newblock \doi{10.18653/v1/D18-2012}.
	\newblock URL \url{https://aclanthology.org/D18-2012}.
	
	\bibitem[Leyton-Brown et~al.(2022)Leyton-Brown, Mausam, Nandwani, Zarkoob,
	Cameron, Newman, Raghu, et~al.]{leyton2022matching}
	Kevin Leyton-Brown, Mausam, Yatin Nandwani, Hedayat Zarkoob, Chris Cameron,
	Neil Newman, Dinesh Raghu, et~al.
	\newblock Matching papers and reviewers at large conferences.
	\newblock \emph{arXiv preprint arXiv:2202.12273}, 2022.
	
	\bibitem[Li et~al.(2015)Li, Wang, Liu, Wang, and Wu]{Li15concert}
	Lei Li, Yan Wang, Guanfeng Liu, Meng Wang, and Xindong Wu.
	\newblock Context-aware reviewer assignment for trust enhanced peer review.
	\newblock \emph{PLOS ONE}, 10\penalty0 (6):\penalty0 1--28, 06 2015.
	\newblock \doi{10.1371/journal.pone.0130493}.
	\newblock URL \url{https://doi.org/10.1371/journal.pone.0130493}.
	
	\bibitem[Liu et~al.(2014)Liu, Suel, and Memon]{liu14graphpropagation}
	Xiang Liu, Torsten Suel, and Nasir Memon.
	\newblock A robust model for paper reviewer assignment.
	\newblock In \emph{Proceedings of the 8th ACM Conference on Recommender
		Systems}, RecSys '14, pp.\  25--32, New York, NY, USA, 2014. ACM.
	\newblock ISBN 978-1-4503-2668-1.
	\newblock \doi{10.1145/2645710.2645749}.
	\newblock URL \url{http://doi.acm.org/10.1145/2645710.2645749}.
	
	\bibitem[Long et~al.(2013)Long, Wong, Peng, and Ye]{Long13gooadandfair}
	Cheng Long, Raymond Wong, Yu~Peng, and Liangliang Ye.
	\newblock On good and fair paper-reviewer assignment.
	\newblock In \emph{Proceedings - IEEE International Conference on Data Mining,
		ICDM}, pp.\  1145--1150, 12 2013.
	\newblock ISBN 978-0-7695-5108-1.
	
	\bibitem[Meir et~al.(2020)Meir, Lang, Lesca, Kaminsky, and
	Mattei]{meir2020market}
	Reshef Meir, J{\'e}r{\^o}me Lang, Julien Lesca, Natan Kaminsky, and Nicholas
	Mattei.
	\newblock A market-inspired bidding scheme for peer review paper assignment.
	\newblock In \emph{Games, Agents, and Incentives Workshop at AAMAS}, 2020.
	
	\bibitem[Merton(1968)]{merton1968matthew}
	Robert~K Merton.
	\newblock The {M}atthew effect in science.
	\newblock \emph{Science}, 159:\penalty0 56--63, 1968.
	
	\bibitem[Mimno \& McCallum(2007)Mimno and McCallum]{mimno07topicbased}
	David Mimno and Andrew McCallum.
	\newblock Expertise modeling for matching papers with reviewers.
	\newblock In \emph{Proceedings of the 13th ACM SIGKDD International Conference
		on Knowledge Discovery and Data Mining}, KDD '07, pp.\  500--509, New York,
	NY, USA, 2007. ACM.
	\newblock ISBN 978-1-59593-609-7.
	\newblock \doi{10.1145/1281192.1281247}.
	\newblock URL \url{http://doi.acm.org/10.1145/1281192.1281247}.
	
	\bibitem[OpenReview(2022)]{OpenReview2022Similarities}
	OpenReview.
	\newblock Paper-reviewer affinity modeling for openreview.
	\newblock \url{https://github.com/openreview/openreview-expertise}, 2022.
	
	\bibitem[Payan(2022)]{payan22fair}
	Justin Payan.
	\newblock Fair allocation problems in reviewer assignment.
	\newblock In \emph{Proceedings of the 21st International Conference on
		Autonomous Agents and Multiagent Systems}, AAMAS '22, pp.\  1857–1859,
	Richland, SC, 2022. International Foundation for Autonomous Agents and
	Multiagent Systems.
	\newblock ISBN 9781450392136.
	
	\bibitem[Peters et~al.(2018)Peters, Neumann, Iyyer, Gardner, Clark, Lee, and
	Zettlemoyer]{peters18deep}
	Matthew~E. Peters, Mark Neumann, Mohit Iyyer, Matt Gardner, Christopher Clark,
	Kenton Lee, and Luke Zettlemoyer.
	\newblock Deep contextualized word representations.
	\newblock In \emph{Proceedings of the 2018 Conference of the North {A}merican
		Chapter of the Association for Computational Linguistics: Human Language
		Technologies, Volume 1 (Long Papers)}, pp.\  2227--2237, New Orleans,
	Louisiana, June 2018. Association for Computational Linguistics.
	\newblock \doi{10.18653/v1/N18-1202}.
	\newblock URL \url{https://aclanthology.org/N18-1202}.
	
	\bibitem[Price et~al.(2010)Price, Flach, and Spiegler]{price2010subsift}
	Simon Price, Peter~A Flach, and Sebastian Spiegler.
	\newblock Subsift: a novel application of the vector space model to support the
	academic research process.
	\newblock In \emph{Proceedings of the First Workshop on Applications of Pattern
		Analysis}, pp.\  20--27. PMLR, 2010.
	
	\bibitem[Rodriguez \& Bollen(2008)Rodriguez and Bollen]{rodriguez08coauthorsip}
	Marko~A. Rodriguez and Johan Bollen.
	\newblock An algorithm to determine peer-reviewers.
	\newblock In \emph{Proceedings of the 17th ACM Conference on Information and
		Knowledge Management}, CIKM '08, pp.\  319--328, New York, NY, USA, 2008.
	ACM.
	\newblock ISBN 978-1-59593-991-3.
	\newblock \doi{10.1145/1458082.1458127}.
	\newblock URL \url{http://doi.acm.org/10.1145/1458082.1458127}.
	
	\bibitem[Shah(2022{\natexlab{a}})]{Shah_2022_randomizedtransparency}
	Nihar~B. Shah.
	\newblock Randomized transparency to mitigate collusions, August
	2022{\natexlab{a}}.
	\newblock URL
	\url{https://researchonresearch.blog/2022/08/19/randomized-transparency-to-mitigate-collusions/}.
	
	\bibitem[Shah(2022{\natexlab{b}})]{shah2021survey}
	Nihar~B Shah.
	\newblock An overview of challenges, experiments, and computational solutions
	in peer review.
	\newblock Communications of the ACM. Extended version at
	\url{https://www.cs.cmu.edu/~nihars/preprints/SurveyPeerReview.pdf},
	2022{\natexlab{b}}.
	
	\bibitem[Shah et~al.(2018)Shah, Tabibian, Muandet, Guyon, and
	Von~Luxburg]{shah2017design}
	Nihar~B Shah, Behzad Tabibian, Krikamol Muandet, Isabelle Guyon, and Ulrike
	Von~Luxburg.
	\newblock Design and analysis of the {NIPS} 2016 review process.
	\newblock \emph{The Journal of Machine Learning Research}, 19\penalty0
	(1):\penalty0 1913--1946, 2018.
	
	\bibitem[Singh et~al.(2022)Singh, D'Arcy, Cohan, Downey, and
	Feldman]{Singh2022SciRepEvalAM}
	Amanpreet Singh, Mike D'Arcy, Arman Cohan, Doug Downey, and Sergey Feldman.
	\newblock Scirepeval: A multi-format benchmark for scientific document
	representations.
	\newblock In \emph{Conference on Empirical Methods in Natural Language
		Processing}, 2022.
	\newblock URL \url{https://api.semanticscholar.org/CorpusID:254018137}.
	
	\bibitem[Squazzoni \& Gandelli(2012)Squazzoni and Gandelli]{squazzoni2012saint}
	Flaminio Squazzoni and Claudio Gandelli.
	\newblock Saint {M}atthew strikes again: An agent-based model of peer review
	and the scientific community structure.
	\newblock \emph{Journal of Informetrics}, 6\penalty0 (2):\penalty0 265--275,
	2012.
	
	\bibitem[Stelmakh et~al.(2021)Stelmakh, Shah, and Singh]{stelmakh21pr4a}
	Ivan Stelmakh, Nihar Shah, and Aarti Singh.
	\newblock {PeerReview4All}: Fair and accurate reviewer assignment in peer
	review.
	\newblock \emph{Journal of Machine Learning Research}, 22\penalty0
	(163):\penalty0 1--66, 2021.
	\newblock URL \url{http://jmlr.org/papers/v22/20-190.html}.
	
	\bibitem[Tan et~al.(2021)Tan, Dai, Ren, Walsh, and Aleksandrov]{tan21envy}
	Mao Tan, Zhuocen Dai, Yuling Ren, Toby Walsh, and Martin Aleksandrov.
	\newblock Minimal-envy conference paper assignment: Formulation and a fast
	iterative algorithm.
	\newblock In \emph{2021 5th Asian Conference on Artificial Intelligence
		Technology (ACAIT)}, pp.\  667--674, 2021.
	\newblock \doi{10.1109/ACAIT53529.2021.9731163}.
	
	\bibitem[Tang et~al.(2010)Tang, Tang, and Tan]{tang10constraied}
	Wenbin Tang, Jie Tang, and Chenhao Tan.
	\newblock Expertise matching via constraint-based optimization.
	\newblock In \emph{Proceedings of the 2010 IEEE/WIC/ACM International
		Conference on Web Intelligence and Intelligent Agent Technology - Volume 01},
	WI-IAT '10, pp.\  34--41, Washington, DC, USA, 2010. IEEE Computer Society.
	\newblock ISBN 978-0-7695-4191-4.
	\newblock \doi{10.1109/WI-IAT.2010.133}.
	\newblock URL \url{http://dx.doi.org/10.1109/WI-IAT.2010.133}.
	
	\bibitem[Thorngate \& Chowdhury(2014)Thorngate and
	Chowdhury]{thorngate2014numbers}
	Warren Thorngate and Wahida Chowdhury.
	\newblock By the numbers: Track record, flawed reviews, journal space, and the
	fate of talented authors.
	\newblock In \emph{Advances in Social Simulation}, pp.\  177--188. Springer,
	2014.
	
	\bibitem[Thurner \& Hanel(2011)Thurner and Hanel]{thurner2011peer}
	Stefan Thurner and Rudolf Hanel.
	\newblock Peer-review in a world with rational scientists: Toward selection of
	the average.
	\newblock \emph{The European Physical Journal B}, 84\penalty0 (4):\penalty0
	707--711, 2011.
	
	\bibitem[Tran et~al.(2017)Tran, Cabanac, and Hubert]{tran17expertsuggestion}
	H.~D. Tran, G.~Cabanac, and G.~Hubert.
	\newblock Expert suggestion for conference program committees.
	\newblock In \emph{2017 11th International Conference on Research Challenges in
		Information Science (RCIS)}, pp.\  221--232, May 2017.
	\newblock \doi{10.1109/RCIS.2017.7956540}.
	
	\bibitem[Triggle \& Triggle(2007)Triggle and Triggle]{triggle07future}
	Chris Triggle and David Triggle.
	\newblock What is the future of peer review? why is there fraud in science? is
	plagiarism out of control? why do scientists do bad things? is it all a case
	of: " all that is necessary for the triumph of evil is that good men do
	nothing?".
	\newblock \emph{Vascular health and risk management}, 3:\penalty0 39--53, 02
	2007.
	
	\bibitem[Wade(2022)]{wade2022semantic}
	Alex~D. Wade.
	\newblock The semantic scholar academic graph ({S2AG}).
	\newblock In \emph{Companion Proceedings of the Web Conference 2022 (WWW’22
		Companion)}, 2022.
	
	\bibitem[Wieting et~al.(2019)Wieting, Gimpel, Neubig, and
	Berg-Kirkpatrick]{wieting-etal-2019-simple}
	John Wieting, Kevin Gimpel, Graham Neubig, and Taylor Berg-Kirkpatrick.
	\newblock Simple and effective paraphrastic similarity from parallel
	translations.
	\newblock In \emph{Proceedings of the 57th Annual Meeting of the Association
		for Computational Linguistics}, pp.\  4602--4608, Florence, Italy, July 2019.
	Association for Computational Linguistics.
	\newblock \doi{10.18653/v1/P19-1453}.
	\newblock URL \url{https://aclanthology.org/P19-1453}.
	
	\bibitem[Wieting et~al.(2022)Wieting, Gimpel, Neubig, and
	Berg-kirkpatrick]{wieting-etal-2022-paraphrastic}
	John Wieting, Kevin Gimpel, Graham Neubig, and Taylor Berg-kirkpatrick.
	\newblock Paraphrastic representations at scale.
	\newblock In \emph{Proceedings of the 2022 Conference on Empirical Methods in
		Natural Language Processing: System Demonstrations}, pp.\  379--388, Abu
	Dhabi, UAE, December 2022. Association for Computational Linguistics.
	\newblock URL \url{https://aclanthology.org/2022.emnlp-demos.38}.
	
	\bibitem[Wu et~al.(2021)Wu, Guo, Wu, Kidambi, Van Der~Maaten, and
	Weinberger]{ruihan21making}
	Ruihan Wu, Chuan Guo, Felix Wu, Rahul Kidambi, Laurens Van Der~Maaten, and
	Kilian Weinberger.
	\newblock Making paper reviewing robust to bid manipulation attacks.
	\newblock In Marina Meila and Tong Zhang (eds.), \emph{Proceedings of the 38th
		International Conference on Machine Learning}, volume 139 of
	\emph{Proceedings of Machine Learning Research}, pp.\  11240--11250. PMLR,
	18--24 Jul 2021.
	\newblock URL \url{https://proceedings.mlr.press/v139/wu21b.html}.
	
	\bibitem[Xu et~al.(2019)Xu, Zhao, Shi, and Shah]{xu2018strategyproof}
	Yichong Xu, Han Zhao, Xiaofei Shi, and Nihar Shah.
	\newblock On strategyproof conference review.
	\newblock In \emph{IJCAI}, 2019.
	
	\bibitem[Zhao et~al.(2022)Zhao, Anand, and Sharma]{zhao2022reviewer}
	Yue Zhao, Ajay Anand, and Gaurav Sharma.
	\newblock Reviewer recommendations using document vector embeddings and a
	publisher database: Implementation and evaluation.
	\newblock \emph{IEEE Access}, 10:\penalty0 21798--21811, 2022.
	
\end{thebibliography}

~\\~\\

\noindent \textbf{{\LARGE Appendices}}

\bigskip

\appendix

\newcommand{\abstr}{a}
\newcommand{\gabstr}{A}
\newcommand{\contrast}{t}
\newcommand{\gabs}{G}

\section{More details on the survey used for data collection}
\label{appendix:survey}

In this section we provide full instructions that were given to the participants of our data-collection procedure.

\medskip

  \begin{mdframed}

  \begin{center}\texttt{\Large{Dataset of Reviewing Expertise}}\end{center}

  \smallskip
  
  \noindent \texttt{The goal of this experiment is to collect a dataset to help researchers design better\\algorithms for computing similarities between papers and reviewers: these algorithms\\will help to improve matching of reviewers to papers in many conferences like NeurIPS, AAAI, ACL, etc.}

  \medskip

  \noindent \texttt{WHAT DO YOU NEED TO DO? \\ ***Recall 5-10 papers in your broad research area that you read to a reasonable extent in the last year and tell us your expertise in reviewing these papers.***\\(HINT: To quickly recall what papers you read recently, you can search for arxiv.org\\or an analogous website in your browser history.)}

  \medskip
  \noindent \texttt{WHICH PAPERS TO REPORT?}
  \begin{itemize}[leftmargin=20pt, itemsep=0pt, topsep=1pt]
      \item \texttt{Papers should not be authored by you}
      \item \texttt{Papers should be freely available online (preferably arXiv, but other open sources are also fine)}
  \end{itemize}

  \medskip

  \noindent \texttt{**Suggestions**}
  \begin{itemize}[leftmargin=20pt, itemsep=0pt, topsep=1pt]
      \item \texttt{Try to choose a set of papers such that some pairs are well-separated and some are very close in terms of your expertise}
      \item \texttt{Please try to avoid ties in the expertise ratings you provide}
      \item \texttt{Try to think of some papers that are less famous to make the dataset more diverse}
      \item \texttt{Try to provide some examples that are not obvious and may be tricky for the \\similarity-computation algorithms. For example, a naive computation of similarity may think that a paper on ``Theory of Information Dissemination in Social \\Networks'' has high similarity with an Information Theory researcher, but in \\reality, this researcher may not have expertise in reviewing this paper}
  \end{itemize}

  \medskip

  \noindent \texttt{WHAT PARTS OF DATA YOU PROVIDE WILL BE RELEASED?\\To facilitate the development of better algorithms for similarity computation (trained to perform well on your data!), we will publicly release data collected in this survey (except email addresses) in a non-anonymized form. Your email will not be released.}

  \medskip

  \noindent \texttt{WHO IS RUNNING THIS SURVEY?\\ The survey is organized by Graham Neubig, Nihar Shah, Ivan Stelmakh (CMU), and John\\Wieting (CMU -> Google Research). Contact Ivan at stiv@cs.cmu.edu if you have any\\questions.
  }

  \medskip

  \noindent \texttt{[...]}

  \medskip

  \noindent \texttt{List up to 10 papers in your broad research area that you read to a reasonable extent\\in the last year and tell us your expertise in reviewing these papers. Please try to\\enter at least 5 papers.}

  \medskip

  \noindent \texttt{Link to Paper 1:\_\_\_\_\_\_\_\_}
  
  \noindent \texttt{Expertise in reviewing Paper 1:}
  \begin{itemize}[leftmargin=20pt, noitemsep, topsep=1pt]
      \item \texttt{1.0 (I am not qualified to review this paper)}
      \item \texttt{1.25}
      \item \texttt{1.5}
      \item \texttt{1.75}
      \item \texttt{2.0 (I can review some aspects of the paper, but can't make a reliable overall\\judgment)}
      \item \texttt{...}
      \item \texttt{3.0 (I can provide an adequate review, but a substantial part of the paper is\\outside of my expertise)}
      \item \texttt{...}
      \item \texttt{4.0 (I have background in most aspects of the paper, but some minor aspects are\\beyond my expertise)}
      \item \texttt{...}
      \item \texttt{5.0 (I have background necessary to evaluate all the aspects of the paper)}
  \end{itemize}

  \medskip

  \noindent \texttt{[...]}

\end{mdframed}

\vspace{5pt}

\section{More details on the \acl{} algorithm}
\label{section:acl}

In this section, we provide more details on the \acl{} algorithm that we evaluate in this paper. 

\paragraph{Training} The model is trained on a large corpus of 45,309 abstracts from the ACL anthology and is inspired by the work of~\citet{wieting-etal-2019-simple,wieting-etal-2022-paraphrastic}. Specifically, the model optimizes a max-margin contrastive learning objective which is defined as follows. First, each abstract $\abstr_i$ from the corpus is split into two disjoint segments of text uniformly at random. These segments are then uniformly at random allocated into two equally-sized groups: $\abstr_i^{(1)} \in \gabstr_1$ and $\abstr_i^{(2)} \in \gabstr_2$.

\smallskip \noindent Second, positive and negative examples are constructed as follows:
\begin{itemize}[itemsep=0pt, leftmargin=20pt, topsep=5pt]
    \item \emph{Positive example:} For each abstract $\abstr_i$, pair ($\abstr_i^{(1)}$, $\abstr_i^{(2)}$) constitutes a positive example.
    
    \item \emph{Negative example:} For each passage $\abstr_{i}^{(1)} \in \gabstr_1$, a counterpart $\contrast_i \ne \abstr_{i}^{(2)}$ from $\gabstr_2$ is selected to maximize the notion of cosine similarity 
    \begin{align*}
        f_{\theta} (\abstr_{i}^{(1)}, t_i) = \cos\left(g(\abstr_{i}^{(1)}, \theta), g(t_i, \theta)\right),
    \end{align*} where $g$ is the sentence encoder with parameters $\theta$. Pair ($\abstr_{i}^{(1)}$, $\contrast_i$) constitutes a negative example.
\end{itemize}

\noindent Finally, with this procedure to build positive and negative examples, the objective of the \acl{} algorithm is defined as:

\begin{align*}
\min_{\theta} \sum_i \max \Big\{\delta - &f_{\theta}(\abstr_{i}^{(1)}, \abstr_{i}^{(2)}) + f_{\theta}(\abstr_{i}^{(1)}, \contrast_{i})), ~0\Big\}.
\end{align*}

\smallskip \noindent Inner-working of the algorithm relies on \texttt{sentencepiece} embeddings~\citep[\url{https://github.com/google/sentencepiece}]{kudo-richardson-2018-sentencepiece} with dimension of 1,024 and vocabulary size of 20,000. The encoder, $g$, simply mean pools the learned \texttt{sentencepiece} embeddings, making for efficient encoding, even on CPU (see~\citealp{wieting-etal-2019-simple,wieting-etal-2022-paraphrastic} for more details on encoding speed). In the training procedure, a batch size of 64 is used and the model is trained for 20 epochs. The margin, $\delta$, is set to 0.4.

\paragraph{Inference} At the inference stage, for a given pair of a submission and a reviewer, the similarity score is defined as a combination of cosine similarities between the submission's abstract and three most-similar abstracts from the reviewer's profile. Specifically, let $\similarity_1, \similarity_2$ and $\similarity_3$ be the top-3 cosine similarities between the submission's abstract and abstracts from the reviewer's profile. The similarity score between the submission and the reviewer is then defined as follows:
\begin{align*}
    \similarity = \similarity_1 + \frac{\similarity_2}{2} + \frac{\similarity_3}{3}.
\end{align*}
\noindent If a reviewer has less than three abstracts in the profile, the corresponding cosine similarity scores $s_i$ are set to be zero.

\section{Additional metrics}
\label{section:metrics}

For completeness, we evaluate the algorithms on several additional metrics. All of these additional evaluations are done on using titles and abstracts of the papers. In addition to the weighted Kendall Tau Distance (WKTD) (\ref{section:setup:metric}), we use WKTD@k, Mean Reciprocal Rank (MRR)@k, Normalized Discounted Cumulative Gain (NDCG)@k, and Precision@k. In order to define these metrics formally, we introduce some notation. The metrics operate separately over each participant in the dataset and over each algorithm, and hence are defined below for any individual participant and any individual algorithm.
\begin{itemize}
    \item The top k papers as ranked by the participant in the dataset is represented as the ordered set $R_k$. The set of all papers for the participant is denoted as $R$.
    \item The set of paper pairs $(p_i, p_j)$ where $p_i\in R_k$ or $p_j\in R_k$ is denoted as $R^2_k$.
    \item The top k papers ranked by the algorithm is represented as the ordered set $A_k$.
    \item The expertise score for a paper $p$ provided by the participant in the dataset is denoted as $\epsilon^{(p)}$. And let $\Delta \epsilon_{ij} = \epsilon^{(p_i)}-\epsilon^{(p_j)}$.
    \item The rank of a paper $p$ given by the algorithm is denoted as $\rho^{(p)}$.
    \item The similarity score determined by an algorithm for a paper $p$ is denoted as $s^{(p)}$. And let $\Delta s_{ij} =s^{(p_i)}-s^{(p_j)}$.
\end{itemize} 
With this notation in place, the additional metrics we evaluate are defined as follows:
\begin{align*}
WKTD@k &=\sum_{(p_i, p_j)\in R^2_k}(I\{\Delta s_{ij}\times \Delta \epsilon_{ij}<0\}\times|\Delta \epsilon_{ij}|+ 
I\{\Delta s_{ij}\times\Delta \epsilon_{ij}=0\}\times \frac{1}{2}|\Delta \epsilon_{ij}|),
\\
MRR@k &= \frac{1}{k}\sum_{p\in R_k}{\rho^{(p)}},\\
NDCG@k &=\frac{DCG@k}{IDCG@k} \text{ where } DCG@k=\sum_{p\in A_k}\frac{\epsilon^{(p)}}{\log_2(i+1)}, \, IDCG@k=\sum_{p \in R_k}\frac{\epsilon^{(p)}}{\log_2(i+1)},\\
Precision@k&=\frac{1}{k}|R_k\cap A_k|.
\end{align*}

For MRR, NDCG, and Precision, we take the mean across all participants for each algorithm as the score. For WKTD, we normalize similar to as described in Section~\ref{section:setup:metric}, where the subscript $r$ represents each participant:
\[WKTD@k_{norm}=\frac{\sum_r WKTD@k_r}{\sum_r\sum_{(p_i, p_j)\in R^2_{r,k}}|\Delta\epsilon_{r,ij}|}\]
We further note that for WKTD, the lower the value the better. For MRR, NDCG, and Precision, the higher the value the better. The results are presented in Tables~\ref{table:kat1}, ~\ref{table:kat3}, ~\ref{table:kat5}, and~\ref{table:kat7}.
\begin{table*}[ht!]
\begin{center}
\begin{small}
\begin{sc}
\begin{tabular}{lcccc}
\toprule

Algorithm &  {Weighted Kendall Tau Distance@1} & {MRR@1} & {NDCG@1} & {Precision@1} \\
\midrule
\tpms{} & $0.29$ & $0.5$ & $0.86$ & $0.24$ \\
\midrule
\elmo{} & $0.34$ & $0.41$ & $0.84$ & $0.17$ \\
\specter{} & $0.23$ & $0.46$ & $0.88$ & $0.24$ \\
\mfr{} & \textbf{0.16} & $0.48$ & \textbf{0.91} & $0.27$ \\
\spectertwo{} Mean Pool & $0.25$ & $0.44$ & $0.87$ & $0.17$ \\
\spectertwo{} 75th Pool & $0.21$ & $0.47$ & $0.88$ & $0.19$ \\
\spectertwo{} Max Pool & $0.17$ & $0.50$ & $0.90$ & $0.33$ \\
\midrule
\acl{} & $0.29$ & $0.48$ & $0.86$ & $0.24$ \\
\midrule
\openai{} & $0.22$ & \textbf{0.55} & $0.89$ & \textbf{0.34} \\
\claude{} & $0.22$ & $0.53$  & $0.89$ & $0.33$ \\
\gemini{} & $0.36$ & $0.46$ & $0.84$ & $0.24$ \\
\bottomrule
\end{tabular}
\end{sc}
\end{small}
\end{center}

\caption{Comparison of similarity-computation algorithms on the collected data using different metrics @top 1. All algorithms operate with reviewer profiles consisting of the 20 most recent papers and use titles and abstracts of papers. Best values are bolded for each column. 
}
\label{table:kat1}
\end{table*}\FloatBarrier

\begin{table*}[ht!]
\begin{center}
\begin{small}
\begin{sc}
\begin{tabular}{lcccc}
\toprule

Algorithm &  {Weighted Kendall Tau Distance@3} & {MRR@3} & {NDCG@3} & {Precision@3} \\
\midrule
\tpms{} & $0.25$ & $0.41$ & $0.89$ & $0.53$ \\
\midrule
\elmo{} & $0.35$ & $0.39$ & $0.85$ & $0.47$ \\
\specter{} & $0.25$ & $0.43$ & $0.89$ & $0.55$ \\
\mfr{} & $0.23$ & $0.44$ & $0.90$ & $0.58$ \\
\spectertwo{} Mean Pool & $0.22$ & $0.43$ & $0.90$ & $0.57$ \\
\spectertwo{} 75th Pool & \textbf{0.19} & $0.44$ & $\textbf{0.91}$ & $0.58$ \\
\spectertwo{} Max Pool & \textbf{0.19} & \textbf{0.46} & $\textbf{0.91}$ & $\textbf{0.59}$ \\
\midrule
\acl{} & $0.26$ & $0.41$ & $0.88$ & $0.54$ \\
\midrule
\openai{} & $0.22$ & $0.44$ & \textbf{0.91} &$0.58$ \\
\claude{} & $0.27$ & $0.43$ & $0.89$ & $0.54$ \\
\gemini{} & $0.37$ & $0.40$ & $0.86$ & $0.48$ \\
\bottomrule
\end{tabular}
\end{sc}
\end{small}
\end{center}

\caption{Comparison of similarity-computation algorithms on the collected data using different metrics @top 3. All algorithms operate with reviewer profiles consisting of the 20 most recent papers and use titles and abstracts of papers. Best values are bolded for each column. 
}
\label{table:kat3}
\end{table*}\FloatBarrier

\begin{table*}[ht!]
\begin{center}
\begin{small}
\begin{sc}
\begin{tabular}{lcccc}
\toprule

Algorithm &  {Weighted Kendall Tau Distance@5} & {MRR@5} & {NDCG@5} & {Precision@5} \\
\midrule
\tpms{} & $0.27$ & $0.38$ & $0.91$ & $0.72$ \\
\midrule
\elmo{} & $0.34$ & $0.37$ & $0.89$ & $0.70$ \\
\specter{} & $0.25$ & $0.39$ & $0.92$ & $0.75$ \\
\mfr{} & $0.22$ & \textbf{0.40} & \textbf{0.93} & \textbf{0.77} \\
\spectertwo{} Mean Pool & $0.23$ & $0.39$ & $0.92$ & $0.76$ \\
\spectertwo{} 75th Pool & $0.22$ & $0.39$ & $0.92$ & $0.75$ \\
\spectertwo{} Max Pool & \textbf{0.20} & \textbf{0.40} & $\textbf{0.93}$ & $\textbf{0.77}$ \\
\midrule
\acl{} & $0.27$ & $0.39$ & $0.91$ & $0.74$ \\
\midrule
\openai{} & $0.23$ & \textbf{0.40} & \textbf{0.93} & $0.76$ \\
\claude{} & $0.29$ & $0.39$ & $0.91$ & $0.72$ \\
\gemini{} & $0.38$ & $0.37$ & $0.89$ & $0.70$ \\
\bottomrule
\end{tabular}
\end{sc}
\end{small}
\end{center}

\caption{Comparison of similarity-computation algorithms on the collected data using different metrics @top 5. All algorithms operate with reviewer profiles consisting of the 20 most recent papers and use titles and abstracts of papers. Best values are bolded for each column. 
}
\label{table:kat5}
\end{table*}\FloatBarrier

\begin{table*}[ht!]
\begin{center}
\begin{small}
\begin{sc}
\begin{tabular}{lcccc}
\toprule

Algorithm &  {Weighted Kendall Tau Distance@7} & {MRR@7} & {NDCG@7} & {Precision@7} \\
\midrule
\tpms{} & $0.28$ & $0.36$ & $0.93$ & $0.81$ \\
\midrule
\elmo{} & $0.35$ & $0.36$ & $0.91$ & $0.80$ \\
\specter{} & $0.27$ & $\textbf{0.37}$ & $0.93$ & $0.81$ \\
\mfr{} & $0.23$ & $\textbf{0.37}$ & $0.94$ & $0.82$ \\
\spectertwo{} Mean Pool & $0.24$ & \textbf{0.37} & $0.94$ & $\textbf{0.83}$ \\
\spectertwo{} 75th Pool & $0.23$ & \textbf{0.37} & $0.94$ & $0.82$ \\
\spectertwo{} Max Pool & \textbf{0.21} & \textbf{0.37} & $\textbf{0.95}$ & $0.82$ \\
\midrule
\acl{} & $0.29$ & $0.36$ & $0.92$ & $0.80$ \\
\midrule
\openai{} & $0.24$ & $\textbf{0.37}$ & $0.94$ & $0.82$ \\
\claude{} & $0.30$ & $\textbf{0.37}$ & $0.93$ & $0.81$ \\
\gemini{} & $0.38$ & $0.36$ & $0.91$ & $0.80$ \\
\bottomrule
\end{tabular}
\end{sc}
\end{small}
\end{center}

\caption{Comparison of similarity-computation algorithms on the collected data using different metrics @top 7. All algorithms operate with reviewer profiles consisting of the 20 most recent papers and use titles and abstracts of papers. Best values are bolded for each column. 
}
\label{table:kat7}
\end{table*}\FloatBarrier

\section{LLM Prompt}
\label{section:prompt}
Here is the prompt we used to evaluate LLMs.
\begin{mdframed}\texttt{
You are an expert in assigning reviewers to papers at a top machine learning \\conference. You want to make sure that the reviewers can give high quality and \\relevant reviews to the paper they are assigned. \\
\\
You are given the following input:\\
- "Reviewer papers", which is a selection of the most recent papers by the reviewer, \\containing the title and abstract of each paper. \\
- "New paper", which contains the title and abstract of a paper that is under \\consideration for assignment. \\
Your job is to assign a score, which is represented by an integer between 0 to 100, \\that evaluates how well the new paper matches with the reviewer's expertise based on \\all of their "Reviewer papers".\\
Consider the following criteria:\\
1. The score should be based on the similarity between the new paper and the reviewer papers.\\
2. The score should be higher if the reviewer has more background knowledge and \\expertise in new paper.\\
3. The score should be lower if the reviewer has less background knowledge and \\expertise in new paper.\\
\\
Please provide the score for the new paper exactly in the following format, where the \\content in <> is what you are generating. Do not provide any additional text and do \\not include the <>:\\
    <score>\\
    Explanation: <briefly explain why you gave <score> >\\
\\
NOTE: \\
- The input is in << >>.\\
- Do no hallucinate in the output and generate information that is not present in the \\reviewer or new papers, make sure to explain why you scored a certain value.\\
\\
<<\\
Input: \\
Reviewer papers:\\
1. Title: *title*\\
Abstract: *abstract*\\
2. Title: Title: *title*\\
Abstract: *abstract*\\
3. ...\\
...\\
20. Title: Title: *title*\\
Abstract: *abstract*\\
=====================\\
New paper:\\
<Title: *title*\\
Abstract: *abstract* >\\
>>\\}
\end{mdframed}

\end{document}